\documentclass[12pt]{article}

\def\Tr{{\rm Tr}}

\begin{document}
  \begin{flushright} \begin{small}
        DTP--MSU/01-17 \\ hep-th/0109151
  \end{small} \end{flushright}
\vspace{.5cm}

\begin{center}
{\Large \bf Inverse dualisation and non-local dualities between
Einstein gravity and supergravities} \vskip.5cm

Chiang-Mei Chen\footnote{Email: cmchen@phys.ntu.edu.tw} \\
{\em Department of Physics, National Taiwan University,
     Taipei 106, Taiwan, R.O.C.,}
\vskip.2cm
Dmitri V. Gal'tsov\footnote{Supported by RFBR; Email:
galtsov@grg.phys.msu.su} and Sergei A.
Sharakin\footnote{Supported by RFBR; Email:
sharakin@grg1.phys.msu.su}\\
{\em Department of Theoretical Physics, Moscow State University,
     119899, Moscow, Russia}
{\small(August 29, 2001)}
\end{center}

\begin{abstract}
We investigate non-local dualities between suitably compactified
higher-dimensional Einstein gravity and supergravities which can be 
revealed if one reinterprets the dualised Kaluza-Klein two-forms in 
$D>4$ as antisymmetric forms belonging to supergravities. We find several
examples of such a correspondence including one between the
six-dimensional Einstein gravity and the four-dimensional
Einstein-Maxwell-dilaton-axion theory (truncated $N=4$
supergravity), and others between the compactified eleven and
ten-dimensional supergravities and the eight or ten-dimensional
pure gravity. The Killing spinor equation of the $D=11$
supergravity is shown to be equivalent to the geometric Killing
spinor equation in the dual gravity. We give several examples
of using new dualities for solution generation and demonstrate
how $p$-branes can be interpreted as non-local duals of pure
gravity solutions. New supersymmetric solutions are presented
including $M2\subset 5$-brane with two rotation parameters.
\end{abstract}


\section{Introduction}
Kaluza-Klein (KK) toroidal reduction is based on a geometric
property, originally discovered in the five-dimensional General
Relativity, to convert the compactified gravitational degrees of
freedom into the scalar-vector theory governed by the dilatonic
Maxwell action. In such a way dynamics of the five-dimensional
gravity with a non-null Killing symmetry is described by the
four-dimensional Einstein-Maxwell-dilaton action with a certain
value of the dilaton coupling constant. In the case of the
higher-dimensional Einstein gravity with $d$ commuting Killing
vectors the number of the KK Maxwell fields in the reduced theory
is $d$, while the dilaton expands into the $d\times d$ matrix of
scalar moduli exhibiting the $GL(d,R)$ global symmetry.
Similarly, dimensional reduction of multidimensional
supergravities gives rise to multiplets of fields transforming
under $E_n$ groups \cite{Ju81}.

In revealing hidden symmetries both in Einstein and supergravity
theories an important step is the dualisation of antisymmetric
form fields. The well-known example is the reduction of
Einstein pure gravity to three dimensions. Trading the KK
two-forms for scalars via dualisation in three dimensions one
discovers the $U$-duality group $SL(d+1,R)$ ($d$ is the total
number of compactified dimensions) instead of the expected
$T$-duality $SL(d,R)$. In four dimensions a $U(1)$
electric-magnetic duality ($S$-duality) also becomes manifest.
$U$-duality of the three-dimensional theory embeds $T$ and $S$
dualities  into a larger group. In supergravities one usually
employs dualisation to reduce the rank of antisymmetric forms.
We refer the reader to an exhaustive discussion of the role and
different ways of implementing the dualisation in dimensionally
reduced multidimensional supergravities
\cite{CrJuLuPo98a,CrJuLuPo98b}.

Dualisation in the opposite sense (producing the form fields of
an increased rank) may also reveal hidden symmetries between
different theories. Following conventions of
\cite{CrJuLuPo98a,CrJuLuPo98b}, we use in such cases the name
``inverse dualisation''. Clearly, the lowest dimension where
inverse dualisation makes sense is five, where a two-form is
dualised to a three form. In $D=4$ the rank of a two-form is
unchanged under dualisation: ``neutral dualisation''.
Dualisation gives rise to several alternative representations
of hidden symmetries which maps the space of solutions on
itself in a non-trivial way. This can be used for solution
generating purposes. For example, dimensional reduction of the
$D=4$ Einstein equations to three dimensions leads to the
$SL(2,R)$ Ehlers group \cite{Eh59}. This symmetry is realized
on the space of two variables one of which is the component of
the four-dimensional metric, and another is the twist
potential: a (pseudo)scalar resulting from the dualisation of
the KK vector. Going down to two dimensions one finds the
infinite-dimensional extension of the Ehlers group to the
Geroch group, but here we are interested rather in its finite
$SL(2,R)$ subgroup. An alternative realization of the same
$SL(2,R)$ symmetry group in two dimensions in known as the
Matzner-Misner group, this latter acts directly on two metric
functions. A map between these different realizations of the
same symmetry (Kramer-Neugebauer map \cite{NeKr69}) plays an
important role in solution generation techniques of general
relativity. Another example, which is closer in spirit to our
present considerations, is the Bonnor map \cite{Bo61} between
the $SL(2,R)$ symmetry groups of the static Einstein-Maxwell
system and the stationary pure gravity (for an extension to
dilaton gravity see \cite{GaGaKe95}).

Here we describe new non-trivial realizations of the ``Bonnor
map'' between different gravity and supergravity theories which
are related to inverse and neutral dualisation of KK two-forms
(some earlier results were presented in
\cite{ChGaMaSh99,ChGaSh00}). Starting with the compactified
pure gravity one can use an inverse dualisation in dimensions
$D\geq 5$ to convert the set of the KK two-forms into higher
rank forms which are then reinterpreted as belonging to some
supergravity theory in another dimension. We are looking for
such compactified supergravities which exhibit the same hidden
symmetries as a suitable compactified Einstein gravity, so the
non-local dualities that we find are essentially of the
Bonnor's type. We will show that these dualities relate the
$p$-branes \cite{St97} to some pure gravity solutions. To avoid
confusion, it is worth noting that our present technique is
essentially different from the earlier known method of
generating supergravity $p$-branes via Harrison transformation
\cite{GaRy98}. This latter is based on $U$-duality arising in
the compactified truncated supergravity action. Here we use a
map between both suitably compactified gravity and supergravity
(generically in different dimensions) which is a Bonnor-type
map between different theories enjoying the same hidden
symmetry.

To be explicit, let us consider the $D$-dimensional action
arising from dimensional reduction of the $D+1$ Einstein theory
in a space-like dimension
\begin{equation}\label{SF}
S_{D} = \int d^D x \sqrt{|g|} \left\{ R(g) -
\frac{\gamma}4(\nabla\psi)^2 - \frac{e^{\gamma\psi}}4 F^2
\right\},
\end{equation}
where
\begin{equation}
\gamma = 1 + \frac1{D-2},
\end{equation}
$F=dA$, and $A,\,\psi$ being the KK vector and scalar
respectively. Instead of treating the two-form $F$ as a vector
field strength, one can dualise it to a $(D-2)$-rank form $H$:
\begin{equation} \label{DF}
F = e^{-\gamma\psi}\,{}^{*_D}H,
\end{equation}
where ${}^{*_D}$ denotes the $D$-dimensional Hodge dual.
Considering $H$ as an exterior derivative of a $(D-3)$-form,
$H=dB$, one can derive the equivalent equations of motion from
the action
\begin{equation} \label{SH}
S'_{D} = \int d^Dx \sqrt{|g|} \left\{ R(g) -
\frac{\gamma}4(\nabla\psi)^2 - \frac{e^{-\gamma\psi}}{2(D-2)!}
H^2 \right\}.
\end{equation}
Continuing the reduction process further one can either generate
new antisymmetric forms in lower dimensions via similar
dualisations, or treat the appearing new KK fields as vectors, so
a variety of alternatives will arise. Note that the
reparameterization of antisymmetric forms usually involves the
appropriate Chern-Simons terms in lower dimensions.

The action (\ref{SH}) is a typical $p$-brane producing action
with a particular value of the dilaton coupling constant. A
crucial point for existence of the Bonnor-type maps we are
looking for is the ``right'' value of the dilaton coupling
constant. We will show that this is indeed the case in a number
of physically interesting situations. New dualities are
essentially non-local, since the relation (\ref{DF}) connecting
variables of two theories is a differential equation for the KK
potential. Therefore, given a $p$-brane solution to (\ref{SH}),
one has to solve the set of differential equations in order to
find the corresponding higher-dimensional metric, and vice
versa. The simplest nontrivial example of such a relationship is
provided by six-dimensional Einstein theory. After
compactification of one dimension and dualisation of the
corresponding KK two-form one obtains five-dimensional gravity
coupled to a dilaton and a three-form. Therefore, the string
solution to the $5D$ gravity with a three-form theory may be
seen as a non-local dual to some pure gravity solution in six
dimensions \cite{ChGaMaSh99}.

The main purpose of the present paper is to describe non-local
dualities involving eleven and ten-diemsnional supergravities.
The plan of the paper is as follows. We start with reviewing the
standard KK dimensional reduction and discuss particular cases
of the resulting four and three-dimensional theories (Sec.
\ref{SKKR}). Then the dimensional reduction of the
six-dimensional relativity to five, four and three dimensions
is considered and an alternative matrix representation of the
$SL(4,R)$ symmetry is derived (Sec. 3). In Sec. 4 we present
the reduction scheme with partial inverse dualisation in
arbitrary dimensions. The detailed description of duality
between the compactified $D=11$ supergravity and the $D=8$ pure
gravity is given in Sec. 5. Then (Sec. 6) several dualities
between IIA and IIB supergravities and ten and
eight-dimensional Einstein gravities are exhibited. The Sec. 7
is devoted to establishing connection between Killing spinor
equation in the compactified $D=11$ supergravity and the
geometric Killing spinor equation in eight dimensions. Several
examples of using new dualities for solution generation are
presented in Sec. 8. We conclude with brief remarks in Sec. 9.
Some mathematical details are given in three Appendices.

\section{The standard KK reduction}\label{SKKR}
Consider the toroidal compactification of the $D+d$ Einstein
gravity to $D$ dimensions starting with the following
parameterization of the metric:
\begin{equation}
ds_{D+d}^2 = g_{dmn}(d\zeta^m + A^m_M dx^M) (d\zeta^n + A^n_N
dx^N) + e^{-\psi/(D-2)} \, ds_{D}^2,
\end{equation}
where
\begin{equation}
g_{d} = ||g_{dmn}||, \qquad  e^\psi = |\det g_{d}|.
\end{equation}
It is assumed that all fields depend only on coordinates in the
$D$-sector. Define a (pseudo)unimodular moduli matrix
\begin{equation}\label{psi}
M_{mn} = g_{dmn} e^{-\psi/d},
\end{equation}
so that $\det M=\epsilon$, with $\epsilon=+1$ ($\epsilon=-1$) if
the metric $g_{d}$ has Euclidean (Lorentzian) signature, or
equivalently, the metric $g_{D}$ has Lorentzian (Euclidean)
signature.\footnote{We consider here only $\Sigma\times K$
space-times with exactly one `time' direction.} Then one obtains
the following action for the reduced theory
\begin{equation}\label{redth}
S_{D} = \int d^Dx \sqrt{|g|} \Bigl\{ R(g) -
\frac{\gamma}4(\nabla\psi)^2 - \frac14 e^{\gamma\psi} F^T M F +
\frac14 g^{MN} \Tr\left(\partial_M M \partial_N M^{-1}\right)
\Bigr\}, \label{SD}
\end{equation}
where
\begin{equation}
\gamma = \frac1d + \frac1{D-2}.
\end{equation}
This action is manifestly invariant under $T$-duality $SL(d,R)
\times R$:
\begin{eqnarray}
&& M \to \Omega^T M \Omega, \quad F \to \Omega^{-1} F,
  \qquad \Omega \in SL(d,R), \label{OMO} \\
&& \psi \to \psi + r, \qquad F \to e^{-\gamma r/2} F,
  \quad r \in R. \label{R}
\end{eqnarray}
The symmetry holds in any dimension $D\geq 2$, and it gets
enhanced in $D\leq 4$. Consider a particular case $D=4$. Then
$\gamma=(2+d)/2d$ and the equations of motion consist of Einstein
equations
\begin{eqnarray}
R_{\mu\nu} &=& \frac{2+d}{8d} \nabla_{\mu}\psi \nabla_{\nu}\psi -
\frac14 \Tr \left(\nabla_{\mu} M \nabla_{\nu} M^{-1}\right)
\nonumber \\
&+& \frac12 e^{(d+2)\psi/2d} \left(F^T_{\mu\alpha} M
F_\nu{}^\alpha - \frac14 g_{\mu\nu} F^T M F \right), \label{FE}
\end{eqnarray}
the dilaton equation
\begin{equation} \label{FP}
\nabla^2 \psi = \frac12 e^{(d+2)\psi/2d} F^T M F,
\end{equation}
Maxwell equations
\begin{equation}\label{FF}
\nabla_\mu \left(e^{(d+2)\psi/2d} M F^{\mu\nu}\right) = 0,
\end{equation}
and an equation for the moduli matrix
\begin{equation}\label{FM}
\nabla(M^{-1}\nabla M) = \frac12 e^{(d+2)\psi/2d} \left( FF^TM -
\frac1d F^TMFI_{d}\right).
\end{equation}
One also has the Bianchi identity
\begin{equation}\label{BF}
\nabla_\mu (\widetilde F^{\mu\nu}) = 0,
\end{equation}
where $\widetilde F$ stands for a four-dimensional dual (
$\widetilde F \equiv {}^{*_4}F$). In addition to continuous
$T$-duality $SL(d,R) \times R = GL(d,R)$ this system is also
symmetric under discrete electric-magnetic duality: the Eqs.
(\ref{FE},\ref{FP},\ref{FM}) remain invariant, while the Maxwell
equation (\ref{FF}) and the Bianchi identity (\ref{BF}) are
interchanged under the discrete $S$-duality transformation
\begin{equation}
\psi \to -\psi, \quad M \to \epsilon M^{-1}, \quad F \to
e^{(d+2)\psi/2d} M \widetilde F.
\end{equation}
Note that the factor $\epsilon$ in the transformation of the
moduli matrix is necessary in view of the relation $\widetilde
F_{\mu\nu} \widetilde F^{\nu\lambda} \equiv -\epsilon F_{\mu\nu}
F^{\nu\lambda}$. In the case $\epsilon=1$ this discrete symmetry
is the symmetry of the initial $(D+d)$-dimensional field
equations, while for $\epsilon=-1$ it relates theories with the
signature $\hbox{diag}(-,+,...,+;+,...,+)$ (only one time-like
direction) and the signature $\hbox{diag}(+,-,...,-;+,...,+)$
($d-1$ time-like directions). About dimensional reduction in
supergravities with non-standard signatures see
\cite{Bar97,Nish98}.

Now let us go down to three dimensions. The resulting action will
read
\begin{equation}
S_{3} = \int d^3x \sqrt{|h|} \Bigl\{ R_{3} - \frac14(\nabla\psi)^2
+ \frac14 \Tr(\nabla g_{d} \nabla g_{d}^{-1}) - \frac14 e^\psi
F^T g_{d} F \Bigr\}.
\end{equation}
It can be transformed into the action for a gravity coupled
sigma-model after dualising the Maxwell two-forms as follows
\begin{equation}
e^{\psi} g_{d} F = {}^{*_3} d \Psi,
\end{equation}
where $\Psi$ is the set of scalars. In terms of new variables the
equations of motion may be obtained from the action
\begin{equation}
S_{3}' = \int d^3x \sqrt{|h|} \Bigl\{ R_3 - \frac14(\nabla\psi)^2
+ \frac14 \Tr(\nabla g_{d} \nabla g_{d}^{-1}) - \frac{\epsilon}2
e^{-\psi} \nabla\Psi^T g_{d}^{-1} \nabla\Psi \Bigr\},
\end{equation}
which simplifies to
\begin{equation} \label{3d}
S_{3}' = \int d^3x \sqrt{|h|} \left\{ R_{3} + \frac14
\Tr(\nabla{\cal M} \nabla{\cal M}^{-1}) \right\},
\end{equation}
where the matrix
\begin{equation} \label{M1}
{\cal M} = \left(\begin{array}{cc}
  e^{-\psi}      & e^{-\psi} \Psi^T \\
  e^{-\psi} \Psi & \epsilon g_{d} + e^{-\psi} \Psi \Psi^T
  \end{array} \right)
\end{equation}
is a symmetric $(d+1)\times(d+1)$ matrix with $\det {\cal M} =
\epsilon^{(d+1)}$. It is easy to see that for $\epsilon=1$,
${\cal M} \in SL(d+1,R)/SO(d+1)$, while for $\epsilon=-1$, ${\cal
M} \in SL(d+1,R)/SO(d-1,2)$. Indeed, the matrix ${\cal M}$ may be
presented in the following vielbien split
\begin{equation}\label{MMN}
{\cal M}_{MN} = {\cal E}^A_M \, \eta_{AB} \, {\cal E}^B_N,
\end{equation}
where
\begin{equation}\label{E1}
{\cal E}^A_M = \left( \begin{array}{cc}
  e^{-\psi/2} & e^{-\psi/2}\Psi_m \\
  0           & \epsilon e^a_m
  \end{array} \right), \quad
\eta_{AB} = \left( \begin{array}{cc}
  1 & 0 \\ 0 & \epsilon\eta_{ab}
  \end{array} \right),
\end{equation}
so that $g_{mn} = e^a_m \eta_{ab} e^b_n$. Therefore for
$\epsilon=1$ one has the Euclidean signature metric $\eta_{AB}$,
while for $\epsilon=-1$ the metric is pseudo-Euclidean. This is
the {\em standard} representation of the three-dimensional
reduction of a higher-dimensional pure gravity \cite{Ma79}.
Finally, further reduced to $D=2$, the action (\ref{3d}) leads
to a completely integrable theory.

\section{Alternative reduction of $D=6$ Einstein gravity and EMDA}
Our first example of using inverse dualisation to relate pure
gravity to (compactified) supergravity starts with six
dimensions. A standard dimensional reduction of $D+d=6$
Einstein theory to five dimensions ($d=1$ in notation of the
previous section) gives the action (\ref{SF}) with $D=5$. Now,
instead of further direct reduction, let us first make an
inverse dualisation of KK two-form to get the five-dimensional
gravity coupled to a dilaton $\hat\phi$ and a three-form field
${\hat H}=d{\hat B}$
\begin{equation}
S_{5} = \int d^5x \sqrt{-\hat g_{5}} \left\{ \hat R_{5} - \frac12
(\partial {\hat\phi})^2 - \frac{e^{-\alpha{\hat\phi}}}{12} {\hat
H}^2 \right\},
\end{equation}
where the dilaton coupling is $\alpha^2=8/3$. This action can be
reinterpreted in supergravity terms. To get further insight into
the nature of the theories involved we perform reduction to four-
and then to the three-dimensional theory. Compactifying first
along a space-like dimension
\begin{equation}
d\hat s_{5}^2 = e^{-4\varphi} \left( d y + {\cal A}_\mu dx^\mu
\right)^2 + e^{2\varphi} ds_{4}^2,
\end{equation}
one obtains for the metric part
\begin{equation}
\sqrt{-\hat g_{5}} \hat R_{5} = \sqrt{-g_{4}} \Big\{ R_{4} - 6
g^{\mu\nu} \partial_\mu \varphi \partial_\nu \varphi - \frac14
e^{-6\varphi} {\cal F}_{\mu\nu} {\cal F}^{\mu\nu} - 2\nabla_\mu
\left( g^{\mu\nu}\partial_\nu \varphi \right) \Big\},
\end{equation}
where ${\cal A} = {\cal A}_\mu d x^\mu,\; {\cal F} = d {\cal A}$,
while the five-dimensional two-form ${\hat B}$ is decomposed into
the four-dimensional two-form $B$ and the one-form $A$
\begin{equation}
{\hat B} = B - d y \wedge A.
\end{equation}
Omitting a total divergence we arrive at the
following four-dimensional action
\begin{equation}
S_{4} = \int d^4x \sqrt{-g_{4}} \Big\{ R_{4} - \frac12(\partial
\phi)^2 - \frac14(\partial \psi)^2 - \frac14 e^{\psi-\phi} {\cal
F}^2 - \frac14 e^{-\psi-\phi} F^2 - \frac{e^{-2\phi}}{12} H^2
\Bigr\},
\end{equation}
where new scalar fields are introduced
\begin{equation}
\phi = \frac12 (\alpha \hat\phi + 4 \varphi), \qquad \psi =
\frac12 (\alpha \hat\phi - 8 \varphi),
\end{equation}
and the four-dimensional strength $H$ will include an appropriate
Chern-Simons term
\begin{equation}
H = d B - {\cal A} \wedge F,
\end{equation}
where $F = d A$. The value of the dilaton coupling constant
corresponding to dimensional reduction of $D=6$ pure gravity
theory $\alpha^2=8/3$ coincides with that arising in toroidal
of the heterotic string effective action. Assuming this value
and dualising the three-form
\begin{equation}
e^{-2\phi} H = -{}^{*_4} d \kappa,
\end{equation}
one can present the corresponding action as
\begin{eqnarray}\label{S4'}
S_{4}' &=& \int d^4x \sqrt{-g} \Big\{ R - \frac12 (\partial
\phi)^2 - \frac14 (\partial \psi)^2 - \frac12 e^{2\phi} (\partial
\kappa)^2 \nonumber \\
&-& \frac14 e^{\psi-\phi}{\cal F}^2 - \frac14 e^{-\psi-\phi}F^2 -
\frac{\kappa}4\left( F{\tilde {\cal F}} + {\cal F}{\tilde F}
\right) \Big\}. \label{S4}
\end{eqnarray}

Now we reduce the theory to three dimensions using the general
partial dualisation procedure described in the Appendix A. With
$d_1=d_2=1$ and $\phi=\phi_1+\phi_2$, our prescription
(\ref{161})-(\ref{168}) leads to the following set of
three-dimensional fields: two electric $v_1, v_2$ and two magnetic
$u_1, u_2$ potentials, scale factor $f$ and twist potential
$\chi$. Together with already introduced scalars we obtain a
three-dimensional $\sigma$-model with the nine-dimensional target
space $\Phi^A = (f, \chi, \phi, \psi, \kappa, v_1, u_1, v_2, u_2)$
endowed with the following metric
\begin{eqnarray}\label{TS4}
dl^2 &=& \frac1{2f^2} \left\{ df^2 + \left[ d\chi + \frac12
\left( v_a du_a - u_a dv_a \right) \right]^2 \right\} + \frac12
d\phi^2 + \frac14 d\psi^2 + \frac12 e^{2\phi} d\kappa^2
\nonumber \\
&-& \frac1{2f} \Big[ e^{\psi-\phi} dv_1^2 + e^{-\psi+\phi} (du_1
- \kappa dv_2)^2 + e^{-\psi-\phi} (dv_2)^2 + e^{\psi+\phi} (du_2
- \kappa dv_1)^2 \Big]. \label{TSM}
\end{eqnarray}
This is the metric of the symmetric space $SL(4,R)/SO(2,2)$ on
which the $SL(4,R)$ isometry group acts transitively. As the
coset representative one can choose a symmetric $SL(4,R)$ matrix,
so that the target space metric will read
\begin{equation}
dl^2 = -\frac14 \Tr \left( d {\cal M} d {\cal M}^{-1} \right).
\end{equation}
The matrix ${\cal M}$ is given by the Eq.(\ref{M2}) with the
following $2\times 2$ real blocks
\begin{eqnarray}
P_1 &=& e^{\psi/2} \left( \begin{array}{cc}
  f e^{-\psi} - (v_1)^2 e^{-\phi} & -v_1 e^{-\phi} \\
  -v_1 e^{-\phi} & -e^{-\phi}
  \end{array} \right), \\
P_2 &=& e^{-\psi/2} \left( \begin{array}{cc}
  f e^\psi - (v_2)^2 e^{-\phi} & -v_2 e^{-\phi} \\
  -v_2 e^{-\phi} & -e^{-\phi}
  \end{array} \right), \\
Q &=& \left( \begin{array}{cc}
  \frac12 \xi-\chi & u_2 - \kappa v_1 \\
  u_1 - \kappa v_2 & -\kappa
  \end{array} \right),
\end{eqnarray}
where
\begin{equation}
\xi = v_1 \left( u_1 - \kappa v_2 \right) + v_2 \left( u_2 -
\kappa v_1 \right).
\end{equation}
Matrix ${\cal M}$ provides an alternative to (\ref{M1})
representation of the coset $SL(4,R)/SO(2,2)$. For $d=3$ the
standard matrix (\ref{M1}) is presented in terms of $3\times 3$,
$1\times 3$ blocks and one scalar, whereas (\ref{M2}) is
parameterized more symmetrically in terms of $2\times 2$ blocks.
New representation provides a direct link to the
three-dimensional reduction of the four-dimensional
Einstein-Maxwell-Dilaton-Axion (EMDA) theory \cite{GaKe96}
(truncated $D=4, N=4$ supergravity with only one vector) in which
case one has a $Sp(4,R)$ matrix. To obtain the latter from the
present model it is enough to identify the electromagnetic
potentials $u_1=u_2,\, v_1=v_2$ and set $\psi=0$ (for more
details see \cite{ClGa01}). Another particular case is the
three-dimensional reduction of the EMDA theory with two vector
fields in which case the relevant symmetry is $SU(2,2)$
\cite{GaSh97} corresponding to the eight-dimensional target
space. Note that both $Sp(4,R)$ and $SU(2,2)$ sigma-models are
K\"ahler, while the present theory is not (its target space is
odd-dimensional).

The existence of two essentially different parameterizations of
the same coset suggests a solution generating technique. Given a
solution in terms of one parameterization, one can construct new
solution performing the identification of variables in terms of
another representation. In the previous paper \cite{ChGaMaSh99}
we obtained a family of five-dimensional string solutions
dualising the five-dimensional   Kerr metric. Here we give
another example taking as a seed the following six-dimensional
pure gravity solution \footnote{The one-center version of this
solution may be considered as plane wave propagating in the
Euclidean Taub-NUT background.}:
\begin{equation}\label{WKN}
ds_{6}^2 = F_2^{-1} ( dz + U du + a_i dx^i)^2 + F_1 du^2 + 2 du
(dv - \omega_i dx^i) + F_2 \delta_{ij} dx^i dx^j,
\end{equation}
with $i,j=1,2,3$, and the harmonic functions
\begin{equation}
\partial^2_{\vec x}F_1 = \partial^2_{\vec x}F_2
= \partial^2_{\vec x}U = 0, \quad \nabla \times \vec \omega =
\nabla U, \quad \nabla \times \vec a = \nabla F_2.
\end{equation}
This metric is a superposition of two five-dimensional pure
gravity solutions: Dobiasch-Maison \cite{DoMa82} (whose
one-center version corresponds to $pp$-wave) and
Gross-Perry-Sorkin \cite{GrPe83,So83} (KK monopole), endowed
with NUT. Three harmonic functions $F_1$, $F_2$, and $U$ on the
three-dimensional transverse space correspond to an electric
charge, a momentum and a NUT charge. From here we can derive
two different $5D$ string counterparts choosing one or another
Killing vector from $\partial_z$, $\partial_u$ for dimensional
reduction from six to five dimensions. Compactifying along
$\partial_z$ we obtain an electrically charged NUT-ed string
superposed with the Brinkmann wave:
\begin{eqnarray}
d \hat s_{5}^2 &=& F_1 F_2^{-1/3} du^2 + 2F_2^{-1/3} du (dv -
\omega_i dx^i)+ F_2^{2/3} \delta_{ij} dx^i dx^j,\\
\hat B &=& F_2^{-1} du \wedge (dv - \omega_i dx^i),\\
e^{\alpha\hat\phi} &=& F_2^{-4/3}. \label{NSW}
\end{eqnarray}
This family of solutions contains an extremal electric string
($U=F_1=1$, $F_2=1+2q/r$, $r^2=\delta_{ij}x^ix^j$) and the pure
gravity Dobiash-Maison solution (at $U=F_2=1$, $F_1=1+2p/r$).
Compactifying along $\partial_u$ one obtains the NUT-ed
magnetically charged $0$-brane superposed with KK monopole along
one of the transversal spatial dimensions:
\begin{eqnarray}
d\hat s_{5}^2 &=& -T^{-2/3} \left\{ dv - \omega_i dx^i + U
F_2^{-1}(dz + a_j dx^j) \right\}^2 \nonumber\\
&+& T^{1/3} \left\{ F_2^{-1} (dz + a_i dx^i)^2 + F_2 \delta_{ij}
dx^i dx^j \right\}, \\
e^{\alpha\hat\phi} &=& T^{4/3}, \label{NMK}
\end{eqnarray}
with the three-form
\begin{eqnarray}
\hat H &=& d (\omega_i dx^i) \wedge (dv - \omega_j dx^j) - d (b_i
dx^i) \wedge (dz + a_j dx^j) \nonumber \\
&+& d \left( U F_2^{-1} (dv - \omega_i dx^i) \wedge (dz + a_j
dx^j) \right),
\end{eqnarray}
where $T=F_1+U^2F_2^{-1}$ and $\nabla \times \vec b = \nabla
F_1$. It reduces to the single extremal $0$-brane if $U=0,
F_2=1, F_1=1+2p/r$, to the pure pure gravity Gross-Perry-Sorkin
monopole for $U=0, F_1=1, F_2=1+2p/r$ and to NUT solution for
$F_1=F_2=1, U\sim n/r$.

Alternatively these solutions can be obtained as {\it null
geodesics} of the target space of the non-linear $\sigma$-model
(\ref{171}).\footnote{For application of null geodesics method
(harmonic maps) in the supergravity context see
\cite{GaRy98,ClGa96}.} To show this it is sufficient to choose as
the geodesic generator a degenerate $sl(4,R)$ matrix of the rank
one and to construct a solution depending on three harmonic
functions. Along these lines one can also construct a four-charge
solution. For an asymptotically flat solution one can define the
asymptotical charges as
\begin{eqnarray}
f &\sim& 1-2M/r, \quad \phi \sim 2D_\phi/r, \quad
\psi \sim 2D_\psi\sqrt{2}/r, \nonumber\\
v_{1,2} &\sim& 2Q_{1,2}/r, \qquad u_{1,2} \sim 2P_{1,2}/r, \\
\chi &\sim& -2N/r, \qquad \kappa \sim 2A/r, \nonumber
\end{eqnarray}
then the null geodesic condition gives the following BPS bound
\begin{equation}
0 = \frac18 \Tr({\cal B}^2) = M^2 + N^2 + A^2 + D_\phi^2 +
D_\psi^2 - Q_1^2 - Q_2^2 - P_1^2 - P_2^2.
\end{equation}
The signature of the target-space metric (\ref{TS4}) tell us that
the most general extremal solution which can be generated via
harmonic maps should contains four independent harmonic
functions, an explicit construction is yet to be done.

\section{Higher rank forms from pure gravity}
Inverse dualisation of a KK two-form in $D$ dimensions gives an
antisymmetric form of the rank $D-2$. With some luck, one can
reinterpret this form as a supergravity matter field thus
providing a non-local duality between gravity and supergravity.
Before passing some realistic possibilities in the context of
string/M theory, here we discuss a modified dimensional
reduction scheme starting with the standard reduction of
$(D+d)$-dimensional pure gravity theory to $D$ dimensions as
given in section \ref{SKKR} and then performing dualisation of
the part $d_1$ of all KK two-forms. It is convenient to
introduce the following notation:
\begin{equation}\label{Fg}
F = dA = \left( \begin{array}{c} F_1 \\ F_2 \end{array} \right),
\quad g_{d} = \left( \begin{array}{cc} g_1 & g_1 K \\
K^T g_1 & g_2 + K^T g_1 K \end{array} \right),
\end{equation}
with $K$ being an arbitrary $d_1 \times d_2$ real matrix, and
$g_1, g_2$ real symmetric $d_1 \times d_1$ and $d_2 \times d_2$
matrices respectively. Rescaling them as
\begin{equation}
M_a = g_a^{-\varepsilon_a} e^{\varepsilon_a \psi_a/d_a},
\end{equation}
where $e^{\psi_a} = \det g_a$, $a=1,2$, $\varepsilon_1 =
-\varepsilon_2 = 1$, so that
\begin{equation}\label{M1M2}
\det M_1 = \det M_2 = 1,
\end{equation}
one obtains
\begin{equation}\label{mod2}
M = e^{-\psi/d} \left(\begin{array}{cc}
  e^{\psi_1/d_1} M_1^{-1}      & e^{\psi_1/d_1} M_1^{-1} K\\
  e^{\psi_1/d_1} K^T M_1^{-1}  &
  e^{-\psi_2/d_2} M_2 + e^{\psi_1/d_1} K^T M_1^{-1} K
  \end{array} \right),
\end{equation}
where $d=d_1+d_2$ and $\psi=\psi_1+\psi_2$. Such a decomposition
of the moduli matrix (\ref{mod2}) and the set of gauge fields can
be viewed as a two-step standard KK reduction (see Appendix B).
In terms of new variables the $D$-dimensional Lagrangian of the
reduced theory (\ref{redth}) can be written as follows
\begin{eqnarray}\label{nond}
{\cal L}_{D} &=& R_{D} - \frac14 \Bigl\{ d_1 (\nabla\phi_1)^2 +
d_2 (\nabla\phi_2)^2 - \frac1{D+d-2}(d_1 \nabla\phi_1 - d_2
\nabla\phi_2)^2 \Bigr\} \nonumber\\
&+& \frac14 \Tr(\nabla M_1 \nabla M_1^{-1}) + \frac14 \Tr(\nabla
M_2 \nabla M_2^{-1}) - \frac12 e^{\phi_1+\phi_2}
\Tr(\nabla K^T M_1^{-1} \nabla K M_2^{-1}) \nonumber\\
&-& \frac14 \Bigl\{ e^{-\phi_2} F_2^T M_2 F_2 + e^{\phi_1}(F_1 +
K F_2)^T M_1^{-1} (F_1 + K F_2) \Bigr\},
\end{eqnarray}
where
\begin{equation}
\phi_a = \varepsilon_a \left( \frac1{d_a} \psi_a + \frac1{D-2}
\psi \right). \label{Fb}
\end{equation}
Now we dualise the part $d_1$ of the KK two-forms $F_{1[2]}$ to
$(D-2)$-forms, $G_{1[D-2]}=dB_{1[D-3]}$:
\begin{equation}
e^{\phi_1} M_1^{-1} (F_1 + K F_2) = {}^{*_D} G_1,
\end{equation}
then an equivalent $D$-dimensional action will read
\begin{eqnarray}\label{DDD}
{\cal L}_{D}' &=& R_{D} - \frac14 \Bigl\{ d_1 (\nabla\phi_1)^2 +
d_2 (\nabla\phi_2)^2 - \frac1{D+d-2}(d_1 \nabla\phi_1 - d_2
\nabla\phi_2)^2 \Bigr\} \nonumber\\
&+& \frac14 \Tr(\nabla M_1 \nabla M_1^{-1}) + \frac14 \Tr(\nabla
M_2 \nabla M_2^{-1}) - \frac12 e^{\phi_1+\phi_2}
\Tr(\nabla K^T M_1^{-1} \nabla K M_2^{-1}) \nonumber\\
&-& \frac14 \Bigl\{ e^{-\phi_2} F_2^T M_2 F_2 + \frac2{(D-2)!}
e^{-\phi_1} G_1^T M_1 G_1 + {}^{*_D} G_1^T K F_2 + F_2^T K^T\,
{}^{*_D} G_1) \Bigr\}.
\end{eqnarray}

An initial non-dualised action (\ref{nond}) is invariant under
global $GL(d,R)$ symmetry (\ref{OMO}), (\ref{R}) which can be
decomposed (see Appendix B (\ref{190})-(\ref{193})) into {\it
central, right} and {\it left} subgroups. The central
transformation $GL(d_1,R)\times GL(d_2,R)$ explicitly reads
\begin{eqnarray}
e^{\varepsilon_a\phi_a} &\to& e^{\varepsilon_a\phi_a}(\det {\cal
G}_a)^{-2\varepsilon_a/d_a}, \label{centr}\\
M_a &\to& {\cal G}_a^T M_a {\cal G}_a (\det {\cal G}_a)^{-2/d_a},\\
F_a &\to& ({\cal G}_a)^{\varepsilon_a} F_a, \\
K &\to& {\cal G}_1^T K{\cal G}_2, \label{cenkon}
\end{eqnarray}
where $a=1,2$, ${\cal G}_a\in GL(d_a,R)$ and $\varepsilon_1=+1$,
$\varepsilon_2=-1$. The right transformation is
\begin{eqnarray}
K &\to& K + {\cal R}, \\
F_1 &\to& F_1 - {\cal R} F_2,
\end{eqnarray}
(other fields inert). The left one in the sector of KK forms reads
\begin{equation}\label{l}
F_2 \to F_2 - {\cal L} F_1,
\end{equation}
with $F_1$ inert, while the scalars undergo rather complicated
transformations which are given in the Appendix B. Here ${\cal
R}$ and ${\cal L}$ --- are arbitrary real $d_1\times d_2$ and
$d_2\times d_1$ parameter matrices respectively. All these
 are symmetries of the action (\ref{nond}).

The same symmetries apply to the dualised theory (\ref{DDD}),
with the difference that only the central subgroup remains an
off-shell symmetry. Central transformations are the same as in
the non-dualised theory with $G_1\to {\cal G}_1^{-1}G_1$
replacing the $F_1$-transformation. Right and left symmetries hold
only for the field equations. This asymmetry is the effect of
omitting a divergence term when going from (\ref{nond}) to
(\ref{DDD}) which is not invariant under left and right
transformations. The right transformation now is
\begin{equation}\label{rd}
K \to K + {\cal R},
\end{equation}
with other fields inert, while the left one is
\begin{eqnarray}
G_1 &\to& G_1 + {\cal L}^T \Lambda_1, \\
F_2 &\to& F_2 + {\cal L} \Lambda_2, \label{ld}
\end{eqnarray}
where $\Lambda_1=K^T G_1 - e^{-\phi_2} M_2\,{}^{*_D} F_2$ and
$\Lambda_2=K F_2 - e^{-\phi_1}M_1\,{}^{*_D} G_1$ so that
$d\Lambda_1=d\Lambda_2=0$ on-shell. Transformations in the
scalar sector are the same as in the non-dualised theory, see
Appendix B.

Now the question is whether one can reinterpret the dualised
action as originating from some supergravity theory with
``matter'' antisymmetric forms. To get link with M-theory, one
has to generate the four-form from the KK two-form. This picks
out six dimensions as the dualisation dimension. One has to
start from some higher-dimensional pure gravity theory on
space-time with Killing symmetries to generate KK two-forms in
six dimensions. For IIB theory the seven-dimensional space-time
is selected as giving the five-form field strength, the
five-diemsnional to get the three-form and so on. One has
analyse suitable compactifications of the eleven and
ten-dimensional supergravities and check whether the dilaton
coupling constants corresponding to reduction to lower
dimensions have the same values as given by the Eq.(\ref{DDD}).

\section{$D=11$ supergravity/$D=8$ Einstein gravity}
Let us consider a particular case of the dual theory (\ref{DDD})
with $d_1=d_2=1$. Then the matrices $M_1$ and $M_2$ trivialize,
and the matrix $K$ reduces to a (pseudo)scalar axion $\kappa$:
\begin{eqnarray}\label{D11}
{\cal L}_{D}^{1,\,1} &=& R_{D} - \frac{e^{2\phi}}2
(\nabla\kappa)^2 - \frac12 (\nabla\phi)^2 -
\frac{D}{8(D-2)}(\nabla\psi)^2 \nonumber\\
&-& \frac{e^{-\phi}}2 \left\{
\frac{e^{\frac{D}{2(D-2)}\psi}}{2!}F_{[2]}^2 +
\frac{e^{-\frac{D}{2(D-2)}\psi}}{(D-2)!} G_{[D-2]}^2 \right\}
\nonumber\\
&-& \frac{\kappa}{2!(D-2)!} E^{\mu_1\mu_2\nu_1...\nu_{D-2}}
F_{\mu_1\mu_2} G_{\nu_1...\nu_{D-2}}.
\end{eqnarray}
Here suitable linear combinations of the scalar fields are
introduced:
\begin{equation}\label{p1}
\phi_1 = \phi + \frac{D}{2(D-2)}\psi, \qquad \phi_2 = \phi -
\frac{D}{2(D-2)}\psi.
\end{equation}
Now we choose $D=6$, then the field $G_{[D-2]}$ is a four-form.
To get a link with eleven-dimensional supergravity we have to
consider the dimensional reduction to six dimensions which
generates the appropriate moduli fields. This can be done as
follows. Starting with the bosonic sector of $D=11$ supergravity
\begin{equation}\label{11d}
S_{11} = \int d^{11} x \sqrt{-\hat G} \left\{ \hat R_{11} -
\frac1{2\times 4!} \hat F_{[4]}^2 \right\} - \frac16 \int \hat
F_{[4]} \wedge \hat F_{[4]} \wedge \hat A_{[3]},
\end{equation}
we assume the following $2+3+6$ decomposition of the metric
(with time in the six-dimensional sector)
\begin{equation}\label{ANZ}
d \hat s_{11}^2 = e^{\varphi_2}(dy_1^2+dy_2^2) + e^{\varphi_3}
(dy_3^2+dy_4^2+dy_5^2) + e^{-(2\varphi_2+3\varphi_3)/4} ds^2_{6},
\end{equation}
and  the potential three-form
\begin{equation}
\hat A_{\mu_1\mu_2\mu_3} = B_{\mu_1\mu_2\mu_3}(x), \qquad \hat
A_{\mu y_1 y_2} = A_{\mu}(x), \qquad \hat A_{y_3y_4y_5} =
\kappa(x).
\end{equation}
The fields $\varphi_2$, $\varphi_3$, $B_{[3]}$, $A_{[1]}$,
$\kappa$ and the six-metric $ds^2_{6}=g_{6\mu\nu}dx^\mu dx^\nu$
depend only on coordinates $x^{\mu}$, parameterizing the
six-space. Note that this class of configurations contains $M2$
(delocalised in three directions) and $M5$-branes as well as
$M2\subset 5$-brane \cite{Lam96,Pa96} and some of intersections
\cite{Co96a,Co96b,Co97} including their non-static
generalizations \cite{ChGaSh00}.

The above ans\"atz leads to the following six-dimensional action:
\begin{eqnarray}
S_{6} &=& \int d^6x \sqrt{|g_{6}|} \Bigl\{ R_{6} -
\frac{e^{2\phi}}2 (\nabla\kappa)^2 - \frac12 (\nabla\phi)^2
- \frac3{16} (\nabla\psi)^2 \nonumber\\
&-& \frac12 e^{-\phi} \left[ \frac1{2!} e^{\frac34 \psi}F^2_{[2]}
+ \frac1{4!} e^{-\frac34 \psi} G^2_{[4]} \right] \Bigr\}
- \int \kappa F_{[2]} \wedge G_{[4]} \nonumber \\
&+& \frac13 \int d \left[ \kappa \left( B_{[3]} \wedge F_{[2]} +
G_{[4]} \wedge A_{[1]} \right) \right],
\end{eqnarray}
where
\begin{equation}\label{vp}
G_{[4]}=dB_{[3]},\quad F_{[2]}=dA_{[1]},\quad
\varphi_2 = \frac13 \phi - \frac12 \psi, \quad \varphi_3 = -
\frac23 \phi.
\end{equation}
and the last term can be omitted as a total divergence. The
Chern-Simons term is non-trivial in this reduction. Comparing
this action with the Eq.(\ref{D11}) for $D=6$ we see that both
theories are precisely the same, including the dilaton
couplings. Recall that (\ref{D11}) was obtained by dimensional
reduction of the eight-dimensional Einstein gravity. Therefore
we find a non-local duality between the $2+3+6$ reduction of the
$D=11$ supergravity and $D=8$ pure gravity.

Now we can formulate the following generating technique for
constructing solutions of the $D=11$ supergravity starting with
some solution of the eight-dimensional Einstein equations
admitting two commuting Killing vectors. First, one has to
choose   coordinates adapted to Killing symmetries (generated by
$\partial_{\zeta^m},\,m=1,2$):
\begin{eqnarray}
ds_{8}^2 &=& g_{mn} \left( d\zeta^m + A^m_{\mu} dx^\mu \right)
\left( d\zeta^n + A^n_{\nu} dx^\nu \right)
+ e^{-\psi/4} g_{\mu\nu} dx^\mu dx^\nu, \nonumber\\
e^{\psi} &=& \det ||g_{mn}||, \label{hab}
\end{eqnarray}
and to perform the following identification of the moduli matrix
in temrs of the supergravity fields
\begin{eqnarray}
g_{mn} &=& e^{\psi/2}
  \left( \begin{array}{cc}
    e^\phi & \kappa e^\phi \\
    \kappa e^\phi & e^{-\phi}+\kappa^2 e^\phi
  \end{array} \right), \nonumber \\
dA^m_{[1]} &=& F^m_{[2]}. \label{8DV}
\end{eqnarray}
Then the four-form $G_{[4]} = dB_{[3]}$ has to be built via
inverse dualisation of the Killing two-forms
\begin{equation}
G_{\alpha_1...\alpha_4} = \frac12 e^{\phi+3\psi/4}
E_{\alpha_1...\alpha_4}{}^{\mu\nu} \left( F^1_{\mu\nu} + \kappa
F^2_{\mu\nu} \right).
\end{equation}
The desired solution of $D=11$ supergravity will read
\begin{equation}\label{aNZ}
d \hat s_{11}^2 = g_2^{1/2} \delta_{ab} dz^a dz^b + g_3^{1/3}
\delta_{ij} dy^i dy^j + (g_2 g_3)^{-1/4} g_{6\mu\nu} dx^\mu
dx^\nu,
\end{equation}
where
\begin{equation}
\ln g_2 = \frac23 \phi - \psi, \qquad \ln g_3 = -2 \phi,
\end{equation}
and the $11D$ four-form $\hat F_{[4]}$ is given by
\begin{equation} \label{fANZ}
\hat F_{[4]} = G_{[4]} + F^2_{[2]} \wedge \epsilon_{[2]} +
d\kappa \wedge \epsilon_{[3]}.
\end{equation}
Some new supergravity solutions constructing along these lines
will be given in the next sections, see also
\cite{ChGaSh00,ChGaSh99,Ch00}.

The $SL(2,R)$ symmetry of the compactified space translate into
the following transformations of the field quantities:
\begin{eqnarray}\label{odin}
z &\to& \frac{az+b}{cz+d}, \qquad z = \kappa+ie^{-\phi}, \quad ad
- bc = 1, \\
\psi &\to& \psi + \hbox{const.}, \\
{\cal F}_{[2]} &\to& (cz+d){\cal F}_{[2]}, \qquad {\cal F}_{[2]}
\equiv e^{\psi/2} F_{[2]} + i e^{-\psi/2}\,{}^{*_6} G_{[4]}.
\label{dva}
\end{eqnarray}
This symmetry unifies into a single multiplet the $M2$, $M5$-
and $M2\subset 5$-branes compatible with the ans\"atz; all of
them have a single pure gravity counterpart: eight-dimensional
plane wave solutions delocalized along one axis:
\begin{equation}
ds_8^2 = F(x) d \zeta_1^2 - 2d\zeta_1 dt + d\zeta_2^2 +
\delta_{ij} dx^idx^j, \qquad i,j = 1,...,5,
\end{equation}
where $\partial_{\vec x}^2 F(x)=0$. More precisely, this
solution corresponds to $M5$-brane (or $M2$-brane after
$\zeta_{1,2} \to \zeta_{2,1}$) and $M2\subset 5$ solution can
be obtained via linear coordinate transformation
\begin{equation}
\zeta_1 \to \zeta_1\cos\xi, \qquad \zeta_2 \to
\frac{\zeta_2}{\cos\xi} + \zeta_1\sin\xi.
\end{equation}

As one straightforward application let us exploit the $SL(2,R)$
symmetry (\ref{centr}-\ref{cenkon}, \ref{rd}-\ref{ld}) to
generate new non-static supergravity solutions. Starting with the
rotating $M5$-brane \cite{CvYo97} (entering into the class of the
above configurations with $F_{[2]}=\kappa=0$ and only $G_{[4]}\ne
0$). Than the left transformation can be used to generate
$F_{[2]}$- and $\kappa$-components of the $M$-brane (see the
Eqs.(\ref{st}-\ref{fn}) in the Appendix B). As a result we obtain
the rotating composite $M2\subset 5$-brane with two rotation
parameters
\begin{eqnarray}
d\hat s_{11}^2 &=& H^{-2/3} H'{}^{1/3} \Bigl\{ - H dt^2 +
\frac{2m}{r^3} f \left( \cosh \delta dt - l_1 \sin^2\theta d\phi_1
- l_2 \cos^2\theta \sin^2\psi d\phi_2 \right)^2 \nonumber\\
&+& dy_1^2 + dy_2^2 \Bigr\} + H^{1/3} H'{}^{-2/3}
\left\{ dy_3^2 + dy_4^2 + dy_5^2\right\} \nonumber \\
&+& H^{1/3} H'{}^{1/3} \left\{ f^{-1} \left[
\left(1+\frac{l_1^2}{r^2}\right)\left(1+\frac{l_2^2}{r^2}\right)
- \frac{2m}{r^3} \right]^{-1} dr^2 + r^2 d\Xi_4^2 \right\}.
\end{eqnarray}
The form field reads
\begin{equation}
\hat F_{[4]} = \sin\zeta \, d A_{[3]} + \cos\zeta \,
{}^{\ast_{11}} d(A_{[3]}\wedge dy_3\wedge dy_4\wedge dy_5) +
\tan\zeta \, d \left(\frac{1-H'}{H'} dy_3 \wedge dy_4 \wedge dy_5
\right).
\end{equation}
In these expressions
\begin{eqnarray}
H &=& 1 + \frac{2m\sinh^2\delta}{r^3} f, \qquad
H' = 1 + \frac{2m \cos^2\zeta \sinh^2\delta}{r^3} f, \\
f^{-1} &=& \Upsilon \left(1+\frac{l_1^2}{r^2}\right)
\left(1+\frac{l_2^2}{r^2}\right), \\
A_{[3]} &=& \frac{1-H^{-1}}{\sinh\delta} \bigl( -\cosh\delta dt +
l_1 \sin^2\theta d\phi_1 + l_2 \cos^2\theta \sin^2\psi d\phi_2
\bigr) \wedge dy_1 \wedge dy_2,
\end{eqnarray}
where
\begin{eqnarray}
\Upsilon &=& \cos^2\theta \cos^2\psi +
\frac{\sin^2\theta}{1+l_1^2/r^2}
+ \frac{\cos^2\theta \sin^2\psi}{1+l_2^2/r^2}, \\
d\Xi_4^2 &=& \left( 1 + \frac{l_1^2 \cos^2\theta}{r^2} +
\frac{l_2^2 \sin^2\theta \sin^2\psi}{r^2} \right) \, d\theta^2 +
\left(1+\frac{l_2^2\cos^2\psi}{r^2}\right)\cos^2\theta \, d\psi^2
\nonumber\\
&-& \frac{2l_2^2\sin\theta\cos\theta\sin\psi\cos\psi}{r^2} \,
d\theta d\psi + \left(1\!+\!\frac{l_1^2}{r^2}\right)\sin^2\theta
\, d\phi_1^2 + \left(1\!+\!\frac{l_2^2}{r^2}\right)\cos^2\theta
\sin^2\psi \,d\phi_2^2. \nonumber
\end{eqnarray}
Electric and magnetic charges for this solution are $Q \propto 2m
\sin\zeta \sinh\delta \cosh\delta$, and $P \propto 2m \cos\zeta
\sinh\delta \cosh\delta$. The rotating $M2\subset 5$ solution
reduces to the rotating $M5$-brane under $\cos\zeta=1$ and
reduces to the two-rotational-parameters $M2$-brane under
$\cos\zeta=0$.

\section{ Ten-dimensional supergravities}
Let us start with the IIB theory. Its bosonic sector contains
the NS-NS fields (metric, B-field and dilaton) and the
Ramond-Ramond potentials: the scalar and the two-form
(combining into $SL(2,R)$ multiplets with the corresponding
NS-NS fields) and the four-form with the self-dual five-form
field strength. Here there are essentially two possibilities:
one is related to seven dimensions as a dualisation platform,
another to five dimensions, this latter possibility works for
the IIA theory as well. On the other hand, we need the third
Killing vector to accommodate for the Ramond-Ramond sector. So
the first option is to consider the ten-dimensional Einstein
gravity admitting three commuting Killing symmetries
\begin{eqnarray}
ds_{10}^2 &=& g_{mn} \left( d\zeta^m + A^m_{\mu} dx^\mu \right)
\left( d\zeta^n + A^n_{\nu} dx^\nu \right) + e^{-\psi/5}
g_{7\mu\nu} dx^\mu dx^\nu, \label{hab10}\\
e^\psi &=& \det ||g_{mn}||, \qquad g_{mn} = g_1^{-1/2}
g_2^{-3/4} \left( \begin{array}{cc} 1 & K^T \\
  K & g_2M+KK^T \end{array} \right), \nonumber
\end{eqnarray}
where the label $m$ takes values $m=\#,1,2$. In the notation of
the Appendix B this corresponds to $D=7,d_1=1,d_2=2$ with the
six-parametric moduli matrix $g_{mn}$ decomposed as $1+2$. In the
matrix notation
\begin{eqnarray}\label{10-2}
||dA^m_{[1]}|| &=& ||F^m_{[2]}|| = ||{\cal F}_{[2]},F^a_{[2]}||^T,
\quad K = ||K_a||, \nonumber \\
M &=& ||M_{ab}|| = \left( \begin{array}{cc}
  e^{-\hat\Phi}+\hat\chi^2e^{\hat\Phi} & e^{\hat\Phi}\hat\chi \\
  e^{\hat\Phi}\hat\chi & e^{\hat\Phi} \end{array} \right),
\end{eqnarray}
so the moduli give two scale factors $g_1, g_2$, a doublet $K_a,
a=1,2$ and a unimodular matrix $M$ parameterized by
$\hat\Phi,\hat\chi$. The latter are obvious candidates for the
dilaton and the R-R scalar of IIB theory, while three one forms
$A^\#_{[1]} \equiv {\cal A}_{[1]}$ and $A^a_{[1]}, a=1,2$ can
be used to generate the five-form and the $SL(2,R)$ doublet of
three-forms.

This construction is dual to the following dimensional
reduction of the IIB theory. The Einstein frame metric splits
as $1+2+7$
\begin{equation}\label{111}
d \hat s_{10}^2 = g_1 dy^2 + g_2^{1/2} \delta_{ab} dz^a dz^b +
(g_1g_2)^{-1/5} g_{7\mu\nu} dx^\mu dx^\nu,
\end{equation}
where the scale factors $g_1,g_2$ and the seven-dimensional
metric are the same as in (\ref{hab10}). The form fields are
constructed from the quantities (\ref{10-2}) via
\begin{eqnarray}\label{222}
\hat F_{[3]}^a &=& F_{[2]}^a \wedge dy + \epsilon^{ab} d K_b
\wedge \epsilon_{[2]}, \nonumber\\
\hat F_{[5]} &=& G_{[5]} + (g_1g_2)^{4/5} \, {}^{*_7} G_{[5]}
\wedge dy \wedge \epsilon_{[2]}, \nonumber\\
G_{[5]} &=& (g_1g_2)^{-4/5} \,{}^{*_7}({\cal F}_{[2]} + K_a
F^a_{[2]}).
\end{eqnarray}
The ans\"atz for $\hat F_{[5]}$ is chosen here in a manifestly
self-dual form
\begin{equation}
\hat F_{[5]} = (1+{}^{*_{10}}) G_{[5]}, \qquad G_{[5]} \equiv
dB_{[4]},
\end{equation}
and one can check that the field equations and the Bianchi
identity for the five-form are equivalent to the pure gravity
Einstein equations for the metric (\ref{hab10}). It is worth
noting that in this case both supergravity  and dual gravity
start from the same dimension ten.

The simplest solutions of the form (\ref{111},\ref{222}) are the
$D3$-brane and $D$-string (the latter being delocalised along
$\{z^a\}$). Our class also includes $D1\subset D3$ solution which
gives rise to $D3-$brane in the NS-NS $B$ field, or, in notation
of \cite{Ts96}, the $NS1+D1+D3$ configurations.

The second possibility is inspired by dimensional reduction from
$D=11$ to IIA supergravity. It is relevant for five-brane
configurations and works (with some alterations) both for IIA and
IIB theories. The Einstein frame supergravity metric is taken as
the $2+3+5$ split
\begin{equation}
d \hat s_{10}^2 = g_2^{1/2} \delta_{ab} dz^a dz^b + g_3^{1/3}
\delta_{ij} dy^i dy^j + (g_2 g_3)^{-1/3} g_{5\mu\nu} dx^\mu
dx^\nu,
\end{equation}
where all functions depend on $x^\mu$. This class of solutions
can be reconstructed together with the appropriate form fields
from the eight-dimensional Ricci-flat metrics admitting three
commuting Killing vectors
\begin{eqnarray}
ds_8^2 &=& g_{mn} \left( d\zeta^m + A^m_\mu dx^\mu \right) \left(
d\zeta^n + A^n_\nu dx^\nu \right) + e^{-\psi/3} g_{5\mu\nu}
dx^\mu dx^\nu, \nonumber\\
e^\psi &=& \det ||g_{mn}||, \label{hab8}
\end{eqnarray}
where now the labeling is $m,n=1,2,\#$ with $a,b=1,2$
corresponding to the first two values the indices $m,n$.
Different identifications of the metric variables as well as
different dualisation prescriptions for three Killing two-forms
associated with $A^m_\mu$ apply to IIA and IIB supergtavities
which we list separately (again denoting $A^{\#}_{[1]}\equiv{\cal
A}_{[1]}$).

\bigskip
\noindent{\bf IIA:} \nopagebreak
\begin{eqnarray}
g_{mn} &=& e^{-\phi} g_2^{-1} g_3^{-2/3}
  \left( \begin{array}{cc} g_2^{1/2}g_3M + KK^T &  K \\
  K^T & 1 \end{array} \right), \nonumber \\
||dA^m_{[1]}|| &=& ||F^m_{[2]}|| = ||F^a_{[2]},{\cal F}_{[2]}||^T,
\quad K = ||K_a||, \nonumber\\
M &=& ||M_{ab}|| =
  \left( \begin{array}{cc} e^{\phi} &  e^{\phi}\kappa \\
  e^{\phi}\kappa & e^{-\phi}+e^{\phi}\kappa^2
  \end{array}\right).
\end{eqnarray}

The dualisation prescription for the four-form $\hat F_{[4]}=
d\hat A_{[3]}-\hat F_{[3]} \wedge \hat A_{[1]}$ and the
three-form $\hat F_{[3]}=d\hat A_{[2]}$ are as follows (in terms
of the first and the third KK two-forms)
\begin{eqnarray}
\hat F_{[4]} &=& -e^{\phi} g_2^{-1/2} g_3^{-1} \,{}^{*_5}
(dK_2-\kappa dK_1) + (F_{[2]}^1 + \kappa F_{[2]}^2) \wedge
\epsilon_{[2]} - dK_1 \wedge \epsilon_{[3]}, \nonumber\\
\hat F_{[3]} &=& -e^{-2\phi}g_2^{-5/3}g_3^{-2/3}\, {}^{*_5} ({\cal
F}_{[2]}+K_a F_{[2]}^a) + d \kappa \wedge \epsilon_{[2]},
\end{eqnarray}
while the two-form $\hat F_{[2]}=d\hat A_{[1]}$ is identified
with the second KK field
\begin{equation}
\hat F_{[2]}=F^2_{[2]}.
\end{equation}
The $10D$ dilaton is given by
\begin{equation}
e^{\hat\Phi} = g_2^{-1}e^{-2\phi}.
\end{equation}

\bigskip
\noindent{\bf IIB:} \nopagebreak
\begin{eqnarray}
g_{mn} &=& g_2^{-1/2} g_3^{-1/3}
  \left( \begin{array}{cc} M^{-1} &  M^{-1}K \\
  M^{-1}K^T & g_2+K^TM^{-1}K \end{array}\right), \nonumber \\
||dA^m_{[1]}|| &=& ||F^m_{[2]}|| = ||F_{[2]a},{\cal F}_{[2]}||^T,
\quad K = ||K_a||, \nonumber \\
M^{-1} &=& ||M^{ab}||.
\end{eqnarray}
To generate the $SL(2,R)$ doublet of three-forms $\hat F_{[3]}^a
= d\hat A_{[2]}^a$ (here $a=1$ and $a=2$ correspond to NS-NS and
R-R fields respectively) and the self-dual five-form
\begin{equation}
\hat F_{[5]} = {}^{*_{10}} \hat F_{[5]} = d\hat A_{[4]} - \frac12
\epsilon_{ab} \hat A^a_{[2]} \wedge \hat F_{[3]}^b
\end{equation}
one has to dualise the KK fields as follows
\begin{eqnarray}
\hat F_{[3]}^a &=& -(g_2g_3)^{-2/3} M^{ab}\, {}^{*_5} \left(
F_{[2]b} + K_b{\cal F}_{[2]} \right) + \epsilon^{ab} dK_b \wedge
\epsilon_{[2]}, \nonumber\\
\hat F_{[5]} &=& {\cal F}_{[2]} \wedge \epsilon_{[3]} -
g_2(g_2g_3)^{-2/3}\, {}^{*_5} {\cal F}_{[2]} \wedge
\epsilon_{[2]},
\end{eqnarray}
(here $\epsilon_{12}=\epsilon^{12}=+1$). In this case the $10D$
dilaton $\hat\Phi$ and the zero-form $\hat\chi$ are expressed
through the $SL(2,R)/SO(2)$ matrix $M_{ab}$ as
\begin{eqnarray}
||M_{ab}|| &=& \left( \begin{array}{cc}
  e^{-\hat\Phi}+\hat\chi^2e^{\hat\Phi} & \hat\chi e^{\hat\Phi} \\
  \hat\chi e^{\hat\Phi} & e^{\hat\Phi} \end{array} \right).
\end{eqnarray}

A class of $8D$ plane wave solutions delocalised in two
directions corresponds to $D0+D2+NS5$-brane in $IIA$ theory and
$NS1+D3+D2$-brane in $IIB$ theory. Hidden supergravity
symmetries unifying these non-marginal solutions in the common
multiplets with marginal ones are dual to linear coordinate
transformations in the compactified subspace
($\zeta^1,\zeta^2,\zeta^\#$) of the corresponding pure gravity
theory.

\section{Killing spinor equations} \label{SPINOR}
In this section we perform dimensional reduction of the $11D$
supergravity Killing spinor equations according to the ans\"atz
(\ref{aNZ},\ref{fANZ}), and try to express the resulting
equations in terms of the dual $8D$ pure gravity. We will find
that these equations coincide exactly with the geometric
Killing spinor equations in eight dimensions, i.e. equations
for covariantly constant spinors. Recall that the existence of
Killing spinors exhibits unbroken supersymmetry of the bosonic
supergravity solutions, so a purely bosonic relationship
between $11D$ supergravity and eight-dimensional pure gravity
gravity admits a fermionic extension. Let us start with the
$11D$ equation for the $32$-component Majorana spinors
$\epsilon_{11}$ expressing the vanishing of the supersymmetry
variation of the gravitino:
\begin{equation}
\hat D_M \epsilon_{11} + \frac1{288} \left( \Gamma_M{}^{NPQR} - 8
\delta_M^N \Gamma^{PQR} \right) \hat F_{NPQR} \epsilon_{11} = 0.
\label{SE}
\end{equation}
In this section and in the Appendix C we use gamma-matrices with
the orthonormal (flat) frame indices as well as the orthonormal
components for all other tensor quantities. As usual, multiindex
matrices ($\Gamma^{PQR}$ etc.) are totally antisymmetric products
of $11D$ gamma matrices.

Recall that the basic BPS $M2,\,M5,$ and $M2\subset 5$-branes
possessing one-half supersymmetry admit Killing spinors with $16$
independent components.

For our three block decomposition of the space-time interval it
is convenient to introduce the corresponding $2\times 3\times 6$
decomposition of the gamma-matrices (more details are given in
the Appendix C):
\begin{equation}
\Gamma_a = \gamma_7 \times \rho_a \times \sigma_0, \qquad
\Gamma_i = \gamma_7 \times \rho_z \times \sigma_i, \qquad
\Gamma_\alpha = \gamma_\alpha \times \rho_0 \times \sigma_0,
\label{Gam}
\end{equation}
where $\rho_a, \sigma_i$ are two sets of the standard $2\times 2$
Pauli matrices, $a=x,y$, $i=x,y,z$, $\rho_0$ and $\sigma_0$ is
the unit matrix. Here $\gamma_\alpha$ are the $6D$ gamma-matrices
in the orthonormal frame: $\{\gamma_\alpha,\gamma_\beta\}=
2\eta_{\alpha\beta}$, while $\gamma_7$ is the chiral operator
$\gamma_7=\gamma_0 \gamma_1 \gamma_2 \gamma_3 \gamma_4 \gamma_5$
so that $\gamma_7^2=1$ and $\{\gamma_7, \gamma_\alpha\} = 0$.
Such a representation corresponds to decomposition of the
$SO(1,10)$ spinor with respect to the subgroup $SO(1,5)\times
SO(2)\times SO(3)$. An arbitrary $11D$ spinor $\epsilon_{11}$
splits into the direct product of four $6D$ spinors according to
the pattern $\bf 32=(8,1^+,2)+(8,1^-,2)$. We write
\begin{equation}\label{SPi}
\epsilon_{11}=(\epsilon_1,\; \epsilon_2,\; \epsilon_3,\;
\epsilon_4)^T,
\end{equation}
where $\epsilon_1,\epsilon_2,\epsilon_3,\epsilon_4$ are
eight-component six-dimensional spinors. The Majorana condition
on $\epsilon_{11}$
\begin{equation}\label{11MA}
\epsilon_{11} = \epsilon_{11}^c = C  \epsilon_{11},
\end{equation}
with the choice of the conjugation matrix $C$ compatible with
(\ref{Gam}) translates into the following conditions on $D=6$
spinors (see Appendix C):
\begin{equation}\label{MAJ}
\epsilon_3 = i P \epsilon_2^*, \qquad \epsilon_4 = i P
\epsilon_1^*, \qquad P=(\tau_z \times \tau_x \times \tau_y).
\end{equation}

Under this decomposition the Eq. (\ref{SE}) splits into the set of
three equations, one of which ($M=\alpha$) is the six-dimensional
differential spinor equation ($D_\alpha$ is the corresponding
spinor covariant derivative):
\begin{eqnarray}\label{3}
&& \Bigl\{ 12 D_\alpha + \gamma_\alpha{}^\beta
\partial_\beta(\phi+\psi/2) + e^{3\psi/8+\phi/2} \gamma_7 \left[
- 2({\bf F}^1_\alpha + \kappa {\bf F}^2_\alpha) + (F^1_\alpha +
\kappa F^2_\alpha) \right] \nonumber\\
&& \qquad + i s_z \left[ e^\phi \gamma_7 (\gamma_\alpha{}^\beta
\partial_\beta \kappa - 2 \partial_\alpha \kappa) +
e^{3\psi/8-\phi/2} \left( {\bf F}^2_\alpha - 2F^2_\alpha \right)
\right] \Bigr\} \, \epsilon_i = 0.
\end{eqnarray}
This is applicable to to all of four six-dimensional spinors
($i=1,2,3,4$) with $s_z=+1$ for $\epsilon_1$ and $\epsilon_3$,
and $s_z=-1$ for $\epsilon_2$ and $\epsilon_4$ (in what follows
we denote these spinors $\epsilon_6$ to distinguish from spinors
in other dimensions). Two other are algebraic equations for
$\epsilon_6$ (with the same agreement for $s_z$) which impose
consistency conditions on the bosonic background:
\begin{equation}\label{1}
\left\{ \not\!\partial (\phi-\psi/2) - i s_z \left( e^\phi
\gamma_7 \not\!\partial \kappa + e^{3\psi/8-\phi/2}{\bf F}^2
\right) \right\} \epsilon_6 = 0,
\end{equation}
and
\begin{equation}
\left\{ \not\!\partial (\phi+\psi/2) + e^{3\psi/8+\phi/2}
\gamma_7 \left( {\bf F}^1 + \kappa {\bf F}^2 \right) - i s_z
e^\phi \gamma_7 \not\!\partial \kappa \right\} \epsilon_6 = 0,
\label{2}
\end{equation}
where
\begin{eqnarray}
\not\!\partial &=& \gamma^\alpha \partial_\alpha = \gamma^\alpha
e^\mu_\alpha \partial_\mu, \qquad {\bf F}^m =
\frac12 F^m_{\alpha\beta} \gamma^{\alpha\beta}, \nonumber\\
F^m_\alpha &=& \frac12[\gamma_\alpha, {\bf F}^m] =
F^m_{\alpha\beta} \gamma^\beta, \qquad {\bf F}^m_\alpha = \frac12
F^m_{\beta\gamma} \gamma_{\alpha}{}^{\beta\gamma}.
\end{eqnarray}

If one multiplies the Eqs. (\ref{1}) and (\ref{2}) by
$\gamma_\alpha$ and makes use of the identity $\gamma_\alpha{\bf
F}^m = {\bf F}^m_\alpha + F^m_\alpha$ one can derive the
following two relations
\begin{equation}
i s_z e^{3\psi/8-\phi/2} \{ {\bf F}^2_\alpha + F^2_\alpha \}
\epsilon_6 = \left\{ \partial_\beta (\phi-\psi/2) + i s_z e^\phi
\gamma_7 \partial_\beta \kappa \right\} \left(
\delta_\alpha^\beta + \gamma_\alpha{}^\beta \right) \epsilon_6,
\end{equation}
\begin{eqnarray}
&& e^{3\psi/8+\phi/2} \gamma_7 \left\{ ({\bf F}^1_\alpha + \kappa
{\bf F}^2_\alpha) + (F^1_\alpha + \kappa F^2_\alpha) \right\}
\epsilon_6 \nonumber\\
&& = \left\{ \partial_\beta (\phi+\psi/2) + i s_z e^\phi \gamma_7
\partial_\beta \kappa \right\} \left( \delta_\alpha^\beta +
\gamma_\alpha{}^\beta \right) \epsilon_6.
\end{eqnarray}
Substituting these into the Eqs. (\ref{3}) we obtain
\begin{eqnarray}\label{3a}
&& \Bigl\{ 4D_\alpha - \partial_\alpha (\phi/3+\psi/2) -
\gamma_\alpha{}^\beta \partial_\beta \psi/4 + e^{3\psi/8+\phi/2}
\gamma_7 (F^1_\alpha + \kappa F^2_\alpha) \nonumber\\
&& \qquad - i s_z \left( e^\phi \gamma_7 \partial_\alpha \kappa +
e^{3\psi/8-\phi/2} F^2_\alpha \right) \Bigr\} \epsilon_6 = 0.
\end{eqnarray}
The set of equations (\ref{1}), (\ref{2}) and (\ref{3a})
represents the $11D$ Killing spinor equations in terms of the
$6D$ quantities. From these, only the last one is a differential
equation on the $6D$ spinors, while two others are constraints on
the $6D$ background and the chirality of $\epsilon_6$.

Now we wish to demonstrate that these equations reduce to the
single equation for covariantly constant spinors in the dual
eight-dimensional empty space-time:
\begin{equation}\label{8dk}
D_M \epsilon_8 = \partial_M \epsilon_8 + \frac14 \hat
\Omega^{AB}{}_M {\cal Y}_{AB} \epsilon_8 = 0,
\end{equation}
in the case of two commuting Killing symmetries (\ref{hab}).
Then the $SO(1,7)$ spinor decomposes as $2\times 6$
\begin{equation}
\epsilon_8 = (\epsilon_+, \epsilon_-)^T,
\end{equation}
where $\epsilon_{+}$ and $\epsilon_{-}$ are again some $6D$
spinors: they correspond to a decomposition over the subgroup
$SO(1,5)\times SO(2) \subset SO(1,7)$, namely $\bf 16=8^++8^-$.
Accordingly, the $8D$ gamma-matrices are presented as
\begin{equation}
{\cal Y}_6 = -\rho_x \times I_8 , \qquad {\cal Y}_7 = - \rho_y
\times \gamma_7, \qquad {\cal Y}_\alpha = \rho_x \times
\gamma_\alpha.
\end{equation}

It is convenient to introduce the one-form representation for the
eight-dimensional metric
\begin{eqnarray}
\hat \theta^m &=& {\cal E}^m_n (d \zeta^n + A^n_\nu dx^\nu) =
\theta^m, \nonumber\\
\hat \theta^\alpha &=& e^{-\psi/8} \theta^\alpha, \qquad
\theta^\alpha = e^\alpha_\nu dx^\nu,
\end{eqnarray}
where
\begin{equation}
{\cal E} = ||{\cal E}^m_n|| = e^{\psi/4} \left(\begin{array}{cc}
  e^{\phi/2} & \kappa e^{\phi/2} \\
  0 & e^{-\phi/2} \end{array} \right).
\end{equation}
Then the spinor connection reads
\begin{eqnarray}
\hat \Omega^{mn} &=& -\partial_\beta {\cal E}^m_p g^{pq} {\cal
E}^n_q \theta^\beta - \frac12 {\cal E}^m_p {\cal E}^n_q
\partial_\beta g^{pq} \theta^\beta, \nonumber\\
\hat \Omega^{m\alpha} &=& -\frac12 e^{\psi/8} {\cal E}^m_p {\cal
E}^n_q \partial^\alpha g^{pq} \theta_n + \frac12
e^{\psi/8} F^{m\alpha\beta} \theta_\beta, \label{con}\\
\hat \Omega^{\alpha\beta} &=& \Omega^{\alpha\beta} +
\frac18(\partial^\alpha \psi \theta^\beta -
\partial^\beta \psi \theta^\alpha) - \frac12
e^{\psi/4} F_n{}^{\alpha\beta} \theta^n, \nonumber
\end{eqnarray}
where $F^m = {\cal E}^m_n F^n, ||F^n||=(F^1_{[2]},F^2_{[2]})^T,
||g_{mn}||={\cal E}^T{\cal E}, ||g^{mn}||=||g_{mn}||^{-1}$. Now
the $8D$ equation (\ref{8dk}) splits into the set of $M=m$
(algebraic) equations, and $M=\mu$ $6D$ spinor equation. The
first two coincide with the equations (\ref{1}) and (\ref{2}),
obtained earlier from the $11D$ theory if we assume $s_z=+1$ for
$\epsilon_6=\epsilon_+$ and $s_z=-1$ for $\epsilon_6=\epsilon_-$.
The six-dimensional $M=\mu$ equation can be also presented in the
form which almost coincides with the $11D$ Killing spinor
equation (\ref{3a}) except for the second term. But this term can
be eliminated via the substitution
\begin{equation}
\epsilon'_6 = e^{-(2\phi+3\psi)/24} \epsilon_6.
\end{equation}
This means that the bosonic correspondence between three-block
truncation of the $11D$ supergravity and eight-dimensional
Einstein gravity extends in a natural way to the supersymmetry
condition: the $11D$ Killing spinor equation is equivalent to
the covariantly constant spinor equation in eight dimensions
(\ref{8dk}).

Therefore any Ricci-flat $8D$ space-time with two commuting
isometries which admits covariantly constant spinors generates
some supersymmetric $11D$ supergravity configuration. It is
worth noting that space-times admitting covariantly constant
spinors are not always Ricci-flat but only Ricci-null (due to
integrability condition for the Eq. (\ref{8dk})), see e.g.
\cite{Fig99}, so one should demand $R_{8MN}=0$ in addition to
the existence of parallel spinors to generate supersymmetric
supergravity solutions from eight-dimensional pure gravity
ones. For {\it static} $8D$ space-times this condition is
fulfilled automatically. The correspondence between eight and
eleven-dimensional Majorana spinors reads explicitly
\begin{equation}\label{expr}
\epsilon_{11} = e^{(2\phi+3\psi)/24} (\epsilon_8,\epsilon_8')^T.
\end{equation}
Here $\epsilon_8$ is a covariantly constant $8D$ spinor (which
has $16$ independent real components as well as the $11D$
Majorana spinor) and $\epsilon_8'$ has to be constructed starting
with $\epsilon_8$ through the Majorana conditions
(\ref{11MA},\ref{MAJ}).

This correspondence may be useful also to simplify the check of
supersymmetry for $11D$ supergravity solutions. To investigate
whether some three-block $11D$ supergravity solution possesses
unbroken supersymmetry it is sufficient to check whether the
corresponding eight-dimensional pure gravity dual admits
covariantly constant spinors. Each (complex) $8D$ covariantly
constant spinor correspond exactly to one Majorana Killing
spinor in $11D$ theory.

Another possible application is a generating technique for
supersymmetric $11D$ backgrounds. Note that all spinor connection
forms (\ref{con}) are invariant under $GL(2,R)$ transformations
(\ref{OMO},\ref{R}) and consequently $6D$ the spinor equations
(\ref{1}),(\ref{2}),(\ref{3a}) are invariant too. Therefore one
can apply the $GL(2,R)$ transformations (\ref{odin})-(\ref{dva})
to the seed $11D$ supersymmetric solutions of the form
(\ref{aNZ}),(\ref{fANZ}) in order to obtain new {\it
supersymmetric} ones.

\section{Solution generation} \label{EXAMPLE}
Here we present some examples of application of the $11D/8D$
duality for solution generating purposes. Let us start with the
five-dimensional generalization of the rotating Dobiasch-Maison
solution \cite{ChGaMaSh99} endowed with NUT and smeared to eight
dimensions as follows
\begin{eqnarray}
ds_{8}^2 &=& T \left( dy + A_t dt + A_\phi d\phi \right)^2 +
dz_1^2 + dz_2^2 + dz_3^2 - \frac{\Delta-a^2\sin^2\theta}{\Sigma T}
\left( dt - \cosh\delta \;\omega d\phi \right)^2 \nonumber \\
&+& \Sigma \left( \frac{d r^2}{\Delta} + d\theta^2 +
\frac{\Delta\sin^2\theta}{\Delta-a^2\sin^2\theta}d\phi^2 \right),
\nonumber \\
A_t &=& \frac{\sinh2\delta \left( mr + an\cos\theta + n^2 \right)}
{\Sigma T}, \nonumber \\
A_\phi &=& -\frac{ 2\sinh\delta \left[ n \Delta\cos\theta +
a\sin^2\theta (mr+n^2) \right]}{\Sigma T}, \label{8DDM}
\end{eqnarray}
where
\begin{eqnarray}
\Delta &:=& r^2 - 2mr + a^2 - n^2, \\
\Sigma &:=& r^2 + (a\cos\theta+n)^2, \\
\omega &:=& \frac{2n\Delta\cos\theta + 2a\sin^2\theta(mr+n^2)}
{a^2\sin^2\theta-\Delta}, \\
T &:=& 1 + 2\sinh^2\delta (mr+an\cos\theta+n^2)\Sigma^{-1}.
\end{eqnarray}
First we have to present the seed solution (\ref{8DDM}) in the
desired form (\ref{hab}) and to read off the variables from
(\ref{8DV}). Here there are two possibilities to identify the
variables according to different ordering of two Killing vectors.
One possibility is
\begin{equation}
\phi = -\frac12 \ln T, \quad \psi = \frac34 \ln T, \quad
A_{[1]}^1 = A_t dt + A_\phi d\phi.
\end{equation}
In this case $G_{[4]}$ vanishes and the scale factors $g_2$ and
$g_3$ are
\begin{equation}
g_2 = T^{-4/3}, \qquad g_3 = T.
\end{equation}
The resulting $11D$ metric and the four-form field read as
follows:
\begin{eqnarray}
ds_{11}^2 &=& T^{-2/3} \left[ -
\frac{\Delta-a^2\sin^2\theta}{\Sigma} \left( dt - \cosh\delta
\;\omega d\phi \right)^2 + dy_1^2 + dy_2^2 \right] \nonumber \\
&+& T^{1/3} \left[ \sum_{k=3}^7 dy_k^2 + \Sigma \left( \frac{d
r^2}{\Delta} + d\theta^2 +
\frac{\Delta\sin^2\theta}{\Delta-a^2\sin^2\theta} d\phi^2
\right)\right], \nonumber \\
\hat A_{t12} &=& A_t, \qquad \hat A_{\phi 12} = A_\phi.
\label{2DDM}
\end{eqnarray}
This is the NUT generalization of the rotating $M2$-brane with
one rotation parameter, which is delocalized on five of eight
transverse coordinates.

An alternative ordering of Killing vectors gives
\begin{equation}
\phi = \frac12 \ln T, \quad \psi = \frac34 \ln T, \quad A_{[1]}^2
= A_t dt + A_\phi d\phi.
\end{equation}
The four-form $G_{[4]}$ is generated via inverse dualisation, the
corresponding three-form potential $B_{[3]}$ has the following
non-zero components
\begin{eqnarray}
B_{tz_2z_3} &=& 2 \sinh\delta \left[ m(a\cos\theta+n)-nr \right]
\Sigma^{-1}, \\
B_{\phi z_2z_3} &=& - \sinh2\delta \Bigl\{ m \cos\theta +
(2n\cos\theta-a\sin^2\theta) \left[ nr-m(a\cos\theta+n)\right]
\Sigma^{-1} \Bigr\}. \nonumber
\end{eqnarray}
Extracting the scale factors  $g_2$ and $g_3$
\begin{equation}
g_2 = T^{-2/3}, \qquad g_3 = T^{-1},
\end{equation}
we finally obtain the following $11D$ supergravity solution:
\begin{eqnarray}
ds_{11}^2 &=& T^{-1/3}
\left[-\frac{\Delta-a^2\sin^2\theta}{\Sigma} \left( dt -
\cosh\delta\;\omega d\phi \right)^2 + \sum_{k=1}^5 dy_k^2 \right]
\nonumber\\
&+& T^{2/3} \left[ dy_6^2 + dy_7^2 + \Sigma \left( \frac{d
r^2}{\Delta} + d\theta^2 + \frac{\Delta\sin^2\theta}{\Delta -
a^2\sin^2\theta}d\phi^2 \right)\right], \nonumber \\
\hat A_{t67} &=& B_{tz_2z_3}, \qquad \hat A_{\phi 67} = B_{\phi
z_2z_3}. \label{5DDM}
\end{eqnarray}
This is the NUT generalization of the rotating $M5$-brane with
one rotation parameter, which is delocalized on two of five
transverse coordinates.

The same solutions can be obtained applying similar procedure to
the five-dimensional rotating NUT-ed Gross-Perry-Sorkin monopole
\cite{ChGaMaSh99} smeared to eight dimensions. Finally, we can
apply the {\em left} transformations (\ref{l1},\ref{l2}) to the
above rotating $M5$-brane with the following parameters:
$r=\cot\zeta, l=-\sin\zeta\cos\zeta$ and $s=2 \ln(\sin\zeta)$.
This leads to NUT-ed rotating composite $M2\subset 5$-brane
(dyon):
\begin{eqnarray}
ds_{11}^2 &=& T^{1/3} T'{}^{1/3} \Bigl\{ T^{-1} \Bigl[ -
\frac{\Delta-a^2\sin^2\theta}{\Sigma} (dt-\cosh\delta\;\omega
d\phi)^2 + dy_1^2 + dy_2^2 \Bigr] \nonumber\\
&+& T'{}^{-1} \left( dy_3^2+dy_4^2+dy_5^2 \right)+dy_6^2+dy_7^2 +
\Sigma \left( \frac{d r^2}{\Delta} + d\theta^2 +
\frac{\Delta\sin^2\theta}{\Delta-a^2\sin^2\theta}d\phi^2
\right)\Bigr\}, \nonumber \\
T' &=& 1 + 2 \cos^2\zeta \sinh^2\delta (mr+an\cos\theta+n^2)
\Sigma^{-1}, \nonumber \\
\hat A_{t12} &=& \sin\zeta \; A_t, \qquad
\hat A_{\phi12} = \sin\zeta \; A_\phi, \nonumber \\
\hat A_{345} &=& \tan\zeta \left( T'{}^{-1} - 1 \right),
\nonumber \\
\hat A_{t67} &=& \cos\zeta \; B_{tz_2z_3}, \qquad \hat A_{\phi67}
= \cos\zeta \; B_{\phi z_2z_3}.\label{25DDM}
\end{eqnarray}
This solution reduces to rotating $M5$-brane if $\cos\zeta=1$ and
to $M2$-brane when $\cos\zeta=0$.

Now let us explore which of these solutions preserves unbroken
supersymmetry. As was shown in the previous section, it is
sufficient to check whether the seed $8D$ metric (\ref{8DDM})
admits covariantly constant spinors. From the Eq. (\ref{8dk}) for
$M=y$ we obtain in the asymptotic region $r\to\infty$
\begin{equation}
\sinh\delta \left\{ m(\sinh\delta\, {\cal Y}_{ry} - \cosh\delta\,
{\cal Y}_{rt}) - n\, {\cal Y}_{\theta\phi}\right\} = 0.
\end{equation}
This implies the following necessary condition
\begin{equation}
(m^2+n^2) \sinh^2\delta = 0.
\end{equation}
So we conclude that $\delta\to \infty$, $m e^{2\delta}$, $n
e^{2\delta}$ and $a$ are finite and $8D$ covariantly constant
spinor takes the form
\begin{equation}
\epsilon_8 = T^{-1/4} e^{-\phi/2} e^{-\varphi {\cal
Y}_{\theta\phi}/2} \epsilon_0, \qquad {\cal Y}_{yt} \, \epsilon_0
= \epsilon_0,
\end{equation}
where $\epsilon_0$ is constant spinor and $\tan\varphi =
r\tan\theta/\sqrt{r^2+a^2}$.

In this case (\ref{2DDM}) takes the form of extremal $M2$-brane
solution with harmonic function $T=1+2q/r, (a=n=0, m\to 0,
m\sinh^2\delta=q)$ presented in polar coordinate system:
\begin{eqnarray}
ds_{11}^2 &=& T^{-2/3}(-dt^2+dy_1^2+dy_2^2) + T^{1/3} \left(
\sum_{k=3}^7 dy_k^2 + dr^2 + d\theta^2 + r^2 \sin^2\theta d\phi^2
\right), \nonumber\\
\hat A_{[3]} &=& (1-T^{-1}) dt \wedge dy_1 \wedge dy_2.
\end{eqnarray}
Other two solutions (\ref{5DDM},\ref{25DDM}) represent in this
limit extremal $M5$- and $M2\subset 5$-brane respectively.

Another interesting example is applying the {\em left}
transformation to the following $M5\bot M5$-brane
\begin{eqnarray}
ds_{11}^2 &=& \left( H_1 H_2 \right)^{2/3} \Bigl\{ (H_1 H_2)^{-1}
\left( -dt^2+dx_1^2+dx_2^2+dx_3^2 \right)
+ H_1^{-1} \left( dy_1^2+dy_2^2 \right) \nonumber\\
&+& H_2^{-1} \left( dz_1^2+dz_2^2 \right) + dr^2 + r^2 d\theta^2
+ r^2\sin^2\theta d\phi^2 \Bigr\}, \nonumber\\
\hat A_{\phi z_1z_2} &=& -2 p_1 \cos\theta, \qquad \hat A_{\phi
y_1y_2} = -2 p_2 \cos\theta,
\end{eqnarray}
where
\begin{equation}
H_1 = 1 + \frac{2p_1}r, \qquad H_2 = 1 + \frac{2p_2}r.
\end{equation}
With the choice of parameters $r=\cot\zeta,
l=-\sin\zeta\cos\zeta$ and $s=2 \ln(\sin\zeta)$, one obtains the
``dyonic intersecting $M$--brane''
\begin{eqnarray}
ds_{11}^2 &=& \left( H_1 H_2 \tilde H \right)^{2/3} \Bigl\{ - (H_1
H_2)^{-1} dt^2 + \tilde H^{-1} \left( dx_1^2+dx_2^2+dx_3^2
\right)\nonumber \\
&+& H_1^{-1} \left( dy_1^2+dy_2^2 \right) + H_2^{-1} \left(
dz_1^2+dz_2^2 \right) + dr^2 + r^2 d\theta^2 + r^2\sin^2\theta
d\phi^2 \Bigr\}, \nonumber\\
\hat A_{ty_1y_2} &=& \frac{2p_1\sin\zeta}r H_1^{-1}, \qquad
\hat A_{\phi z_1z_2} = -2 p_1 \cos\zeta \cos\theta, \nonumber\\
\hat A_{tz_1z_2} &=& \frac{2p_2\sin\zeta}r H_2^{-1}, \qquad
\hat A_{\phi y_1y_2} = -2 p_2 \cos\zeta \cos\theta, \nonumber\\
\hat A_{x_1x_2x_3} &=& \tan\zeta \left( 1-\tilde H^{-1} \right),
\end{eqnarray}
where
\begin{equation}
\tilde H = \sin^2\zeta + \cos^2\zeta H_1 H_2.
\end{equation}
The same solution can be obtained by applying the same
transformation to the $M2\bot M2$-brane. For $\sin\zeta=0$ the
solution reduces to $M2\bot M2$-brane and for $\cos\zeta=0$
reduces to $M5\bot M5$-brane. This solution was found previously
in \cite{Co97}.

\section{Discussion}
Our aim was to investigate whether one can find new links
between gravity and supergravities in higher dimensions using
inverse dualisation to generate antisymmetric forms from
Kaluza-Klein vectors. Apparently this is an impossible task for
the full non-abridged supergravities, but as we have shown, it
works for certain consistent truncations. While it is perhaps
not surprising that the Maxwell equations and the Bianchi
identities for the KK fields translate into similar equations
for dual higher rank forms, a non-trivial test is whether the
dilatonic exponents in the reduced actions are the same. We
have found that it is so in several cases. The most interesting
is the correspondence between $2+3+6$ dimensional reduction of
the eleven-dimensional supergravity and eight-dimensional
Einstein gravity with two commuting Killing vectors. A related
duality holds between both (suitably compactified) IIA and IIB
ten-dimensional supergravities and eight-dimensional Einstein
gravity with three commuting Killing vectors. Another case is
the correspondence between the ten-dimensional Einstein gravity
and a suitably compactified IIB theory. It is worth noting that
all dualities of this sort are non-local in the sense that
variables of one theory are related to  variables of the dual
theory not algebraically, but via solving differential
equations.

A remarkable fact is that the $11D$-supergravity/$8D$-gravity
duality holds not only in the bosonic sector, but also extends
to Killing spinor equations exhibiting unbroken supersymmetries
of the $11D$ theory. Namely, the existence of Killing spinors
in the supergravity framework is equivalent to the existence of
covariantly constant spinors in the dual Einstein gravity. It
would be interesting to check whether this correspondence found
at the linearized level extends non-linearly, i.e. holds for
suitably supersymmetrized $8D$ gravity. A more challenging
question is whether classical dualities found here have
something to do with quantum theories. Although we were not
able to give an answer here, our results concerning the
ten-dimensional supergravities look promising in this direction.

As a direct application one can use the above dualities  for
solution generation purposes similarly to the Bonnor map in
general relativity. Fortunately, the truncations considered
leave enough space for $p$-brane solutions including dyonic
states, rotating and NUT-ed configurations, as well as various
brane intersections. We were able to find hitherto unknown
branes (including supersymmetric ones) applying new dualities
for generating classical solutions. Recently there was an
upsurge of interest to  {\em fluxbranes} or $Fp$-branes, which
are the generalizations of the magnetic fluxtube (Melvin
solution). It is worse noting that the first genuine fluxbrane
solution with higher rank antisymmetric forms was constructed
by a method similar to the present one \cite{GaRy98}. The method
described here was also used to  construct  intersecting
fluxbranes  \cite{ChGaSh99}   one of which  was later
rediscovered  by Russo and Tseytlin \cite{RuTs01,Ts01}.

\section*{Acknowledgments}
The work of CMC was supported in part by the Taiwan CosPA
project. This work was also supported in part by the RFBR grant
00-02-16306.

\begin{appendix}

\section{Alternative reduction to $D=4$ and $D=3$ (partial dualisation)}
Here we give details of the alternative reduction scheme based
on partial dualisation of KK two-forms to $D=4$ and $3$. Let us
start with the standard dimensional reduction to four
dimensions, using decompositions (\ref{Fg})-(\ref{Fb}) with
$\gamma=(2+d)/2d$:
\begin{eqnarray}
{\cal L}_{4} &=& R_{4} - \frac14 \Bigl\{ d_1(\nabla\phi_1)^2 +
d_2(\nabla\phi_2)^2 - \frac1{d+2}(d_1\nabla\phi_1 -
d_2\nabla\phi_2)^2 \Bigr\} \nonumber\\
&+& \frac14 \Tr(\nabla M_1\nabla M_1^{-1}) + \frac14 \Tr(\nabla
M_2 \nabla M_2^{-1}) - \frac12 e^{\phi_1+\phi_2} \Tr(\nabla K^T
M_1^{-1}\nabla K M_2^{-1}) \nonumber\\
&-& \frac14 \Bigl\{e^{-\phi_2}F^T_2 M_2 F_2 + e^{\phi_1}(F_1 + K
F_2)^T M_1^{-1}(F_1 + K F_2) \Bigr\}.
\end{eqnarray}
Now we wish to dualise a part $d_1$ of the full set of $d$
Maxwell two-forms as follows ($\tilde F\equiv {}^{*_4}F$)
\begin{equation}
e^{\phi_1} M_1^{-1}(F_1 + K F_2) = \tilde G_1,
\end{equation}
leaving the remaining $d_2$ forms unchanged. The equivalent (dual)
action involves $d_2$ KK potentials $A_{[1]}^{m_2}$ and $d_1$
dual potentials $B_{[1]m_1}$, $dB=G_{[2]}$:
\begin{eqnarray}\label{L4'}
{\cal L}'_{4} &=& R_{4} - \frac14 \Bigl\{ d_1 (\nabla\phi_1)^2 +
d_2(\nabla\phi_2)^2 - \frac1{d+2}(d_1\nabla\phi_1 -
d_2\nabla\phi_2)^2 \Bigr\} \nonumber\\
&+& \frac14 \Tr(\nabla M_1 \nabla M_1^{-1}) + \frac14 \Tr(\nabla
M_2 \nabla M_2^{-1}) - \frac12 e^{\phi_1+\phi_2} \Tr(\nabla K^T
M_1^{-1} \nabla K M_2^{-1}) \nonumber\\
&-& \frac14 \Bigl\{ e^{-\phi_2} F^T_2 M_2 F_2 + e^{-\phi_1} G_1^T
M_1 G_1 + \tilde G_1^T K F_2 + F_2^T K^T \tilde G_1 \Bigr\}.
\end{eqnarray}
Note the presence of Chern-Simons densities in the action. Let us
discuss symmetries of the transformed action. In deriving this
expression the total divergence was omitted which is not
invariant under right and left subgroups of the full $GL(d,R)$
symmetry group. Therefore, only the central subgroup
$GL(d_1,R)\times GL(d_2,R)$ will be the symmetry of the action,
the rest of the group being the symmetry of the equations of
motion.

Now compactify the dualised action (\ref{L4'}) further to three
dimensions. One can expect to have the $\sigma$-model possessing
an enhanced symmetry $SL(d+2,R)$ (here $d=d_1+d_2$ refers to
four-dimensional theory) as before. Let us write the four-metric
as
\begin{equation}\label{161}
ds^2_{4} = g_{\mu\nu} dx^{\mu} dx^{\nu} = - f (dt-\varpi_i
dx^i)^2 + f^{-1} h_{ij} dx^i dx^j,
\end{equation}
and introduce the columns of electric and magnetic potentials
\begin{eqnarray}
G_{1it} &=& \partial_i v_1, \qquad F_{2it} = \partial_i v_2, \\
M_1 G_1^{ij} &=& \frac{-f e^{\phi_1}}{\sqrt{h}}
\epsilon^{ijk}( \partial_k u_1 - K \partial_k v_2), \\
M_2 F_2^{ij} &=& \frac{-f e^{\phi_2}}{\sqrt{h}} \epsilon^{ijk}(
\partial_k u_2 - K^T \partial_k v_1).
\end{eqnarray}
Define the twist potential $\chi$ via
\begin{equation}
\tau_i = \partial_i \chi + \frac12 v_a^T \partial_i u_a -
\frac12 u_a^T \partial_i v_a, \quad a = 1,2,
\end{equation}
where $\tau$ is
dual to the $3D$ KK vector $\varpi_i$
\begin{equation}\label{168}
\tau^i = \frac{f^2}{\sqrt{h}} \epsilon^{ijk} \partial_j \varpi_k
\end{equation}
in accordance with the part of Einstein equations
\begin{equation}
R_{4t}{}^i = \frac{f}{2\sqrt{h}} \epsilon^{ijk} \partial_j
\tau_k.
\end{equation}
As a result we arrive at the $\sigma$-model
\begin{equation}\label{171}
S_3 = \int d^3x \sqrt{h} \left\{ R_3 + \frac14
\Tr\left(\nabla{\cal M} \nabla{\cal M}^{-1} \right) \right\},
\end{equation}
with the following target space metric
\begin{eqnarray}\label{Target}
dl_{\sigma}^2 &=& -\frac14\left( d {\cal M} d {\cal M}^{-1}
\right) \\
&=& \frac{d f^2 + d\tilde \chi^2}{2 f^2} + \frac14 \Bigl\{ d_1 d
\phi_1^2 + d_2 d \phi_2^2 - \frac1{2+d}(d_1 d \phi_1 - d_2 d
\phi_2)^2 \Bigr\} \nonumber\\
&-& \frac14 \Tr(d M_1 d M_1^{-1}) - \frac14 \Tr(d M_2 d M_2^{-1})
+ \frac12 e^{\phi_1+\phi_2}\Tr(d K^TM_1^{-1} d K M_2^{-1})
\nonumber\\
&-& \frac1{2f} \Bigl\{ e^{-\phi_1} dv_1^T M_1 dv_1 + e^{-\phi_2}
dv_2^T M_2 dv_2 + e^{\phi_1} dw_1^T M_1^{-1} dw_1 + e^{\phi_2}
dw_2^T M_2^{-1} dw_2 \Bigr\}, \nonumber
\end{eqnarray}
where
\begin{eqnarray}
d \tilde \chi &=& d \chi + \frac12 v_a^T d u_a - \frac12 u_a^T d
v_a, \\
d w_1 &=&  d u_1 - K d v_2, \\
d w_2 &=& d u_2 - K^T d v_1.
\end{eqnarray}
In deriving this expression we made use of the following
$(d_1+1)\times (d_2+1)$ split of the coset matrix
\begin{eqnarray}\label{M2}
{\cal M} &=&
  \left(\begin{array}{cc} 1 & 0 \\ Q^T & 1 \end{array} \right)
  \left(\begin{array}{cc} P_1^{-1} & 0 \\ 0 & P_2 \end{array}\right)
  \left(\begin{array}{cc} 1 & Q \\ 0 & 1 \end{array} \right)
  \nonumber \\
&=& \left( \begin{array}{cc}
  P_1^{-1}     & P_1^{-1} Q \\
  Q^T P_1^{-1} & P_2 + Q^T P_1^{-1} Q
  \end{array} \right),
\end{eqnarray}
where the $P_1, P_2$ are the real $d_1 \times d_1$ and $d_2\times
d_2$ symmetric matrices with the same determinant, and $Q$ is the
real $d_1\times d_2$ matrix. The trace decomposes as follows
\begin{equation}
\Tr \left( d {\cal M} d {\cal M}^{-1} \right) \equiv \Tr \left(
dP_1 dP_1^{-1} + dP_2 dP_2^{-1} - 2 dQ P_2^{-1} dQ^T P_1^{-1}
\right),
\end{equation}
while the matrices $P_1, P_2$ and $Q$ in terms of previously
introduced variables read
\begin{eqnarray}
P_1 &=& e^{\chi_1} \left( \begin{array}{cc}
  f e^{\phi_1} - v_1^T M_1 v_1 & -v_1^T M_1 \\
  -M_1 v_1                     & -M_1
  \end{array} \right), \nonumber\\
P_2 &=& e^{\chi_2} \left( \begin{array}{cc}
  f e^{\phi_2} - v_2^T M_2 v_2 & -v_2^T M_2 \\
  -M_2 v_2                     & -M_2
  \end{array} \right), \nonumber\\
Q &=& \left( \begin{array}{cc}
  \frac12 \xi - \chi & \varphi_2^T \\
  \varphi_1          & -K
  \end{array} \right),
\end{eqnarray}
where
\begin{eqnarray}
\xi &=& v_a^T \varphi_a, \qquad  a=1,2, \\
\varphi_1 &=& u_1 - K v_2, \quad \varphi_2 = u_2 - K^T v_1, \\
\chi_a &=& -\phi_a + \frac1{d+2} (d_1 \phi_1 - d_2 \phi_2).
\end{eqnarray}
It is easy to see that the target space (\ref{Target}) is a
symmetric space $SL(d+2,R)/SO(d,2)$ as in the standard reduction
scheme. But the representation of the coset matrices now is
given in terms the blocks $(d_1+1)\times (d_1+1)$,
$(d_2+1)\times (d_2+1)$, $(d_1+1)\times (d_2+1)$ and
$(d_2+1)\times (d_1+1)$ with arbitrary $0\leq d_1, d_2\leq d$,
instead of the block structure $1\times (d+1)$ in the standard
scheme (\ref{M1}).

\section{Two-step reduction from $D+d_1+d_2$ to $D$}
Our split (\ref{Fg}) may be regarded as a two-step dimensional
reduction: first from $(D+d_1+d_2)$ dimensions to $(D+d_2)$,
second to final $D$. This corresponds to the following
parameterization of the metric
\begin{eqnarray}
ds_{(D+d)}^2 &=& g_{m_1n_1} (dz^{m_1}+K^{m_1}_{m_2}dy^{m_2}+{\cal
A}^{m_1}_{\mu}dx^{\mu}) (dz^{n_1}+K^{n_1}_{n_2}dy^{n_2}+{\cal
A}^{n_1}_{\nu}dx^{\nu}) \nonumber \\
&+& g_{m_2n_2}(dy^{m_2}+{\cal B}^{m_2}_{\mu}dx^{\mu})
(dy^{n_2}+{\cal B}^{n_2}_{\nu}dx^{\nu}) +
e^{-\psi/(D-2)}g_{D\mu\nu}dx^{\mu}dx^{\nu},
\end{eqnarray}
where $m_1, n_1 = 1,...,d_1$, $m_2, n_2 = 1,...,d_2$ and $\mu,
\nu = 1,...,D$. Introducing matrices
\begin{eqnarray}
g_1 &=& ||g_{m_1n_1}||, \quad g_2 = ||g_{m_2n_2}||, \quad
g_D = ||g_{D\mu\nu}||, \\
K &=& ||K^{m_1}_{m_2}||, \quad {\cal A} = ||{\cal
A}^{m_1}_{\mu}||, \quad {\cal B} = ||{\cal B}^{m_2}_{\mu}||,
\end{eqnarray}
one can write
\begin{equation}
g_{D+d} = ||g_{MN}|| = \left( \begin{array}{ccc}
  g_1          & g_1 K           & g_1{\cal A} \\
  K^T g_1      & g_2 + K^T g_1 K & g_2{\cal B}+K^T g_1{\cal A}\\
  {\cal A}^T g_1 & {\cal B}^T g_2 + {\cal A}^T g_1 K
    & e^{-\psi/2} g_D + {\cal A}^T g_1{\cal A}
      + {\cal B}^T g_2{\cal B}
  \end{array} \right),
\end{equation}
\begin{equation}
F = dA = d \left( \begin{array}{c}
  {\cal A} - K{\cal B} \\
  {\cal B}
  \end{array} \right).
\end{equation}

Consider the action of the $SL(d,R)\times R$  symmetry on this
representation. First we combine (\ref{OMO}, \ref{R}) into the
single $GL(d,R)$ transformation
\begin{equation}\label{190}
{\cal M} \to G^T {\cal M} G, \qquad F \to G^{-1} F,
\end{equation}
where
\begin{eqnarray}
{\cal M} &=& e^{\gamma\psi} M \in GL(d,R)/SO(d), \nonumber \\
G &=& e^{\gamma r/2} \Omega \in GL(d,R).
\end{eqnarray}
For $G$ it is convenient to use the Gauss decomposition into the
product of {\it left} (triangle), {\it center} (block-diagonal)
and {\it right} (triangle) matrices
\begin{equation}\label{193}
G = \left( \begin{array}{cc}
  1 & 0 \\ {\cal L} & 1 \end{array} \right)
  \left(\begin{array}{cc}
  {\cal G}_1^{-1} & 0 \\ 0 & {\cal G}_2 \end{array} \right)
  \left(\begin{array}{cc}
  1 & {\cal R} \\ 0 & 1 \end{array} \right)
\end{equation}
Then we obtain a natural decomposition of the $GL(d,R)$ symmetry
into the following three subgroup:

\smallskip
\noindent {\it central}: see (\ref{centr}-\ref{cenkon}),
$GL(d_1,R)\times GL(d_2,R)$;

\smallskip
\noindent {\it right}: $K \to K + {\cal R}$, $F_1 \to F_1 -
{\cal R} F_2$, other fields inert;

\smallskip
\noindent {\it left}: $F_2 \to F_2 - {\cal L} F_1$, $F_1$ being
inert and
\begin{eqnarray}\label{st}
&& e^{-\phi_2} M_2 + e^{\phi_1} K^T M_1^{-1} K \;\; \hbox{invariant},\\
&& e^{-\phi_1} M_1 + e^{\phi_2} K M_2^{-1} K^T \;\; \hbox{invariant},\\
&& e^{\phi_1} M_1^{-1} \to e^{\phi_1} (1 + K{\cal L})^T M_1^{-1}
(1 + K{\cal L}) + e^{-\phi_2} {\cal L}^T M_2 {\cal L}, \\
&& e^{\phi_1} M_1^{-1} K \to e^{\phi_1}(1 + K{\cal L})^T M_1^{-1}
K + e^{-\phi_2} {\cal L}^T M_2. \label{fn}
\end{eqnarray}
In the dual terms the symmetries are decomposed in a similar way:

\smallskip
\noindent {\it central} (dual): $G_1 \to {\cal G}_1^{-1} G_1$,
other variables as in (\ref{centr}-\ref{cenkon});

\smallskip
\noindent {\it right} (dual): $K \to K + {\cal R}$, other
quantities inert;

\smallskip
\noindent {\it left} (dual): scalar sector transforms as in
(\ref{st})-(\ref{fn}) and
\begin{eqnarray}
G_1 &\to& (1 + K{\cal L})^T G_1 - e^{-\phi_2} {\cal L}^T M_2
\tilde F_2,\\
F_2 &\to& (1 + {\cal L}K) F_2 - e^{-\phi_1} {\cal L} M_1 \tilde
G_1.
\end{eqnarray}
The Bianchi identities $dG_1 = dF_2 = 0$ hold on shell only.
When $d_1=d_2=1$, $M_1$ and $M_2$ trivialize  and the left
(dual) transformations reduce to
\begin{eqnarray}
\kappa &\to& {\cal D} \left[ (1+\kappa l) \kappa + l e^{-2\phi}
\right], \\
e^{-\phi_1} &\to& {\cal D} e^{-\phi_1}, \\
e^{-\phi_2} &\to& {\cal D} e^{-\phi_2},
\end{eqnarray}
where ${\cal D}^{-1}=(1+\kappa l)^2+l^2e^{-2\phi}$,
$2\phi=\phi_1+\phi_2$. This can be rewritten as
\begin{eqnarray}
z^{-1} &\to& z^{-1} + l, \qquad
z = \kappa + i e^{-\phi}, \nonumber \\
\frac32 \psi &=& \phi_1 - \phi_2  \;\;\hbox{invariant}.
\label{l1}
\end{eqnarray}
Gauge fields transform as follows:
\begin{eqnarray}
G_{[4]} &\to& (1 + \kappa l) G_{[4]} - e^{-\phi_2}l
\tilde F_{[2]}, \nonumber \\
F_{[2]} &\to& (1 + \kappa l) F_{[2]} - e^{-\phi_1}l \tilde
G_{[4]}. \label{l2}
\end{eqnarray}

\section{Majorana spinor conditions}
Four six-dimensional spinors which represent a $11D$ spinor are
not independent because of the Majorana condition on
$\epsilon_{11}$ (\ref{11MA}). To find the explicit relations we
choose the following representation for $6D$ gamma-matrices
$\gamma_\alpha$ in terms of the Pauli matrices ($\tau_x, \tau_y,
\tau_z$):
\begin{eqnarray}
\gamma_0 &=& i(\tau_x \times \tau_0 \times \tau_0), \qquad
\gamma_1 = (\tau_y \times \tau_0 \times \tau_0), \qquad \gamma_2
= (\tau_z \times \tau_x \times \tau_0), \nonumber\\
\gamma_3 &=& (\tau_z \times \tau_y \times \tau_0), \qquad
\gamma_4 = (\tau_z \times \tau_z \times \tau_x), \qquad \gamma_5
= (\tau_z \times \tau_z \times \tau_y),
\end{eqnarray}
so that $\gamma_7=(\tau_z \times \tau_z \times \tau_z)$. In this
representation the $11D$ gamma matrices read
\begin{eqnarray}
\Gamma_\alpha &=& \gamma_\alpha \times \rho_0 \times \sigma_0,
\qquad \Gamma_6 = \gamma_7 \times \rho_x \times \sigma_0, \qquad
\Gamma_7 = \gamma_7 \times \rho_y \times \sigma_0, \nonumber\\
\Gamma_8 &=& \gamma_7 \times \rho_z \times \sigma_x, \qquad
\Gamma_9 = \gamma_7 \times \rho_z \times \sigma_y, \qquad
\Gamma_{10} = \gamma_7 \times \rho_z \times \sigma_z,
\end{eqnarray}
and obey the identities
\begin{equation}
\Gamma_0^+ = -\Gamma_0, \qquad \Gamma_M^+ = \Gamma_M, \qquad
\Gamma_M^T = (-1)^{M+1} \Gamma_M.
\end{equation}
The charge conjugation matrix $C$ satisfying
\begin{equation}
C^{-1} \Gamma_M C = -\Gamma_M^T, \qquad C^T = -C, \qquad C^+ C=I,
\end{equation}
can be chosen as
\begin{equation}
C = -\Gamma_0 \Gamma_2 \Gamma_4 \Gamma_6 \Gamma_8 \Gamma_{10} =
(\tau_y \times \tau_x \times \tau_y) \times \rho_x \times
\sigma_y.
\end{equation}
Then the Majorana condition $\epsilon=\epsilon^c=C\Gamma_0^T
\epsilon^j$ translates into the following equation on the $6D$
spinors $\epsilon_1, \epsilon_2, \epsilon_3, \epsilon_4$:
\begin{equation}
\left( \begin{array}{c} \epsilon_3 \\ \epsilon_4 \end{array}
\right) = i \rho_x \times P \left( \begin{array}{c} \epsilon_1 \\
\epsilon_2 \end{array} \right)^*
\end{equation}
where $P = (\tau_z \times \tau_x \times \tau_y)$. Therefore any
$11D$ Majorana Killing spinor $\epsilon_{11}$ may be expressed
through exactly one $8D$ covariantly constant spinor
$\epsilon_{8}=(\epsilon_1,\epsilon_2)^T$ via (\ref{expr}).

\end{appendix}

\end{document}